# Shocks and instabilities in the partially ionised solar atmosphere

Andrew Hillier *, Ben Snow

*Department of Mathematics, University of Exeter, North Park Road, Exeter EX4 4QE, UK*

**Abstract**

The low solar atmosphere is composed of mostly neutral particles, but the importance of the magnetic field for understanding observed dynamics means that interactions between charged and neutral particles play a very important role in controlling the macroscopic fluid motions. As the exchange of momentum between fluids, essential for the neutral fluid to effectively feel the Lorentz force, is through collisional interactions, the relative timescale of these interactions to the dynamic timescale determines whether a single-fluid model or, when the dynamic frequency is higher, the more detailed two-fluid model is the more appropriate. However, as many MHD phenomena fundamentally contain multi-time-scale processes, even large-scale, long-timescale motions can have an important physical contribution from two-fluid processes. In this review we will focus on two-fluid models, looking in detail at two areas where the multi-time-scale nature of the solar atmosphere means that two-fluid physics can easily develop: shock-waves and instabilities. We then connect these ideas to observations attempting to diagnose two-fluid behaviour in the solar atmosphere, suggesting some ways forward to bring observations and simulations closer together.

*Keywords:* Magnetohydrodynamics; Shocks; Instabilities; plasma; Partial ionisation

## 1. The partially ionised solar atmosphere

Observations of the cool material in the lower solar atmosphere show a wide range of both interesting (from a perspective of theoretically understanding them) and important (in terms of their role in mass transfer and heating of the solar atmosphere) dynamic phenomena. At the lowest level of the solar atmosphere, the solar photosphere, observations show cell-like structures known as granules. These are turbulent convective cells carrying heat from the solar interior into the atmosphere. As we move up into the solar chromosphere we find a change in the dynamics as we move from the fluid dominated layers into those where the magnetic field dominates, characterised by the plasma $\beta$ (ratio of gas to magnetic pressure) dropping below 1 (Gary, 2001). Observations show many dynamic features in this layer of the atmosphere including jets (e.g. Nishizuka et al., 2011; Singh et al., 2011) and spicules (e.g. Pereira et al., 2014), which are driven by or excite a wide range of physical processes including waves (e.g. Morton et al., 2014; Okamoto and De Pontieu, 2011), shocks (e.g. Houston et al., 2018; Houston et al., 2020) and magnetic reconnection (e.g. Nishizuka et al., 2008). We can find chromospheric material even high up in the solar atmosphere, suspended in the solar corona as prominences/filaments. Observations of these fascinating structures reveal waves (e.g. Okamoto et al., 2007; Hillier et al., 2013), instabilities (Berger et al., 2008; Berger et al., 2010; Berger et al., 2011; Berger et al., 2017; Hillier and Polito, 2018), reconnection (Hillier and Polito, 2021) and turbulence (Leonardis et al., 2012; Hillier et al., 2017). Overall there is a wide range of dynamic phenomena, all connecting with fundamental MHD theoretical concepts. An example of a prominence and many of the chromospheric dynamics discussed above is presented in Fig. 1.

* Corresponding author.
 E-mail address: a.s.hillier@exeter.ac.uk (A. Hillier).

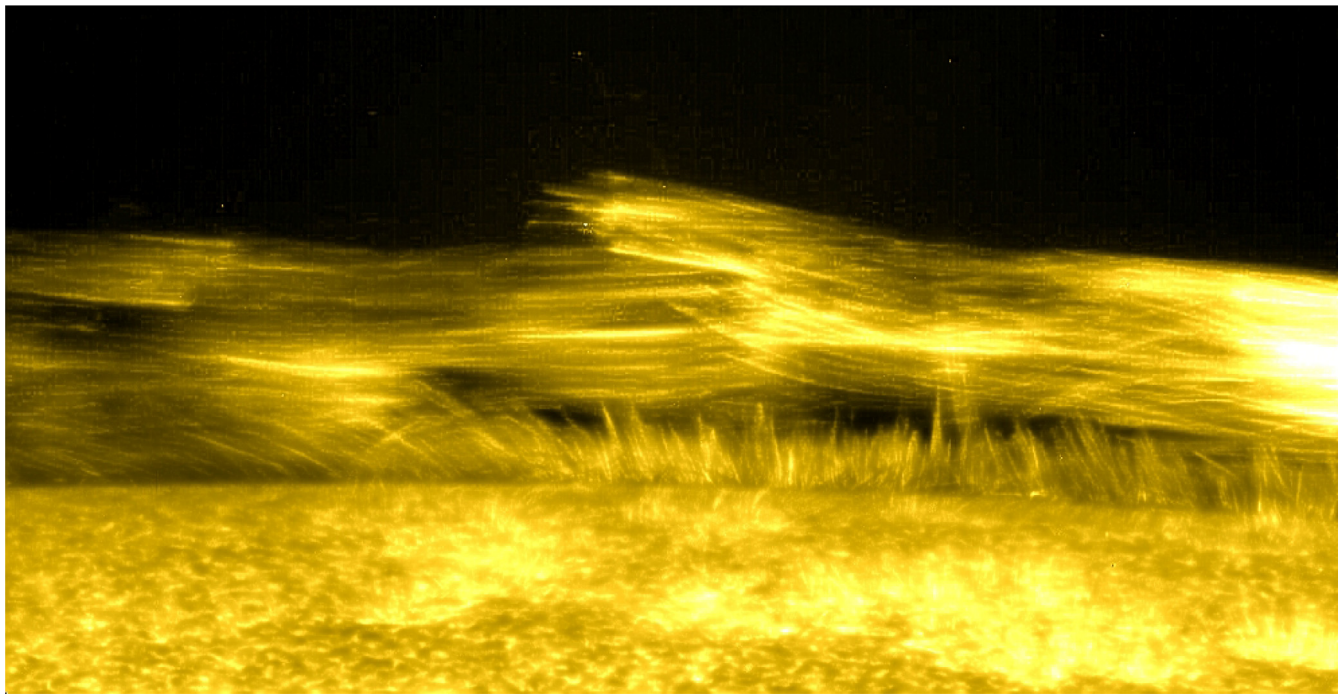

Fig. 1. Image of the solar chromosphere, spicules and a solar prominence observed by Hinode SOT. Image courtesy of J. Okamoto (NAOJ). The dynamics of this prominence are analysed in Okamoto et al., (2016).

There are multiple aspects that physically connect the different dynamics observed in these different layers of the solar atmosphere. The similar temperature of the material, in the range of a few thousand to $10^4$ K, allows all these layers to be treated as the cool phase of the solar atmosphere. And being cool has a very important consequence. The (relatively) low temperatures of the plasma found in these layers of the atmosphere, along with the local densities, is insufficient to fully ionise the material. This results in the plasma in these layers being classed as partially ionised plasma, as a significant proportion of the species that compose it are neutral (Khomenko et al., 2014a).

Another key aspect that connects these dynamics is the importance of magnetic fields for either driving the dynamics, e.g. in launching chromospheric jets, or as a conduit for energy to be transported from lower levels to the atmosphere to higher regions. However, neutral species don't naturally feel the Lorentz force, so the fact that these lower regions of the solar atmosphere are partially ionised means that there has to be a physical process coupling the neutral material to the magnetic field. We will present how this physically happens, and its consequences for some dynamics observed in the solar atmosphere in the subsequent sections.

In this review article, we will present some fundamental ideas behind the modelling of partially ionised plasmas, and that allow us to understand where this will lead to differences in dynamics from a fully ionised MHD approximation. We will focus purely on advances in two-fluid modelling (see the next section for more details) with a particular focus on multi-scale dynamics. The particular examples of this we focus on are shocks (see Section 3) and and instabilities (see Section 4).

## 2. Dynamics in partially ionised plasmas

With the plasma found in the lower solar atmosphere being partially ionised, and the connection between this plasma and the magnetic field of high importance to understand a myriad of observed phenomena, the question is: how does a plasma that is predominantly neutral couple with the magnetic field. For the solar atmosphere, the answer to this question comes through collision-like interactions between charged species (which feel the Lorentz force of the magnetic field) and the neutral particles. These might be hard-sphere collisions or through charge exchange (e.g. Vranjes and Krstic, 2013; Meier and Shumlak, 2012). There are other processes that couple the fluids, with ionisation and recombination an important one for solar plasma due to the role they have in determining the ionisation degree of the different regions of the solar atmosphere. However, the strongest coupling comes through collision processes (e.g. Vranjes and Krstic, 2013; Nóbrega-Siverio et al., 2020a).

There are many ways that partially ionised plasmas have been modelled, from kinetic models following the evolution of individual particles (Jara-Almonte et al., 2019) to single fluid approximations (e.g. Hillier et al., 2010; Cheung and Cameron, 2012; Khomenko et al., 2014b; Nóbrega-Siverio et al., 2020b; Nóbrega-Siverio et al., 2020a; González-Morales et al., 2020) based on ambipolar diffu-sion (Braginskii, 1965). As the charged plasma component of a partially ionised plasma would feel fundamentally different forces than that of the neutral fluid (i.e. whether the fluid feels the Lorentz force or not), one sensible way to model this system is to look at a two-fluid system where the charged species are modelled as a charge-neutral plasma and the neutral species as a separate fluid with the two fluids coupled through collisions and ionisation/recombination (e.g. Khomenko, 2020). The equations for this approximation of the system (assuming a purely hydrogen plasma and including only the collisional coupling terms) are given as

$$\frac{\partial \rho_n}{\partial t} + \nabla \cdot (\rho_n \mathbf{v}_n) = 0, \tag{1}$$

$$\frac{\partial}{\partial t}(\rho_n \mathbf{v}_n) + \nabla \cdot (\rho_n \mathbf{v}_n \mathbf{v}_n + P_n \mathbf{I}) - \rho_n \mathbf{g} = -\alpha_c \rho_n \rho_p (\mathbf{v}_n - \mathbf{v}_p), \tag{2}$$

$$\frac{\partial e_n}{\partial t} + \nabla \cdot [\mathbf{v}_n (e_n + P_n)] - \rho_n \mathbf{v}_n \cdot \mathbf{g} = -\alpha_c \rho_n \rho_p \left[\frac{1}{2}(\mathbf{v}_n^2 - \mathbf{v}_p^2) + \frac{1}{\gamma-1}\left(\frac{P_n}{\rho_n} - \frac{1}{2}\frac{P_p}{\rho_p}\right)\right], \tag{3}$$

$$e_n = \frac{P_n}{\gamma - 1} + \frac{1}{2}\rho_n v_n^2, \tag{4}$$

for the neutral fluid and

$$\frac{\partial \rho_p}{\partial t} + \nabla \cdot (\rho_p \mathbf{v}_p) = 0, \tag{5}$$

$$\frac{\partial}{\partial t}(\rho_p \mathbf{v}_p) + \nabla \cdot \left(\rho_p \mathbf{v}_p \mathbf{v}_p + P_p \mathbf{I} - \mathbf{B}\mathbf{B} + \frac{\mathbf{B}^2}{2}\mathbf{I}\right) - \rho_p \mathbf{g} = \alpha_c \rho_n \rho_p (\mathbf{v}_n - \mathbf{v}_p), \tag{6}$$

$$\frac{\partial}{\partial t}\left(e_p + \frac{\mathbf{B}^2}{2}\right) + \nabla \cdot [\mathbf{v}_p(e_p + P_p) - (\mathbf{v}_p \times \mathbf{B}) \times \mathbf{B}] - \rho_p \mathbf{v}_p \cdot \mathbf{g} = \alpha_c \rho_n \rho_p \left[\frac{1}{2}(\mathbf{v}_n^2 - \mathbf{v}_p^2) + \frac{1}{\gamma-1}\left(\frac{P_n}{\rho_n} - \frac{1}{2}\frac{P_p}{\rho_p}\right)\right], \tag{7}$$

$$\frac{\partial \mathbf{B}}{\partial t} - \nabla \times (\mathbf{v}_p \times \mathbf{B}) = 0, \tag{8}$$

$$e_p = \frac{P_p}{\gamma - 1} + \frac{1}{2}\rho_p v_p^2, \tag{9}$$

$$\nabla \cdot \mathbf{B} = 0, \tag{10}$$

for the plasma. The variables $\rho, P, \mathbf{B}, \mathbf{v},$ and $e$ are the density, pressure, magnetic field, velocity field, and internal energy respectively, and $\gamma$ is the adiabatic index. Note we have performed the simplification of $\mathbf{B}/\sqrt{\mu_0} \rightarrow \mathbf{B}$ in these equations. Here the subscripts $n$ and $p$ refer to the neutral and plasma fluid respectively, and $\mathbf{g}$ is the gravitational acceleration. We have assumed that these are ideal gases with $P_n = R_g \rho_n T_n$ and $P_p = 2R_g \rho_p T_p$, with $R_g$ the gas constant and $T_n$ and $T_p$ the neutral and plasma temperatures. General information of derivation of such equations can be found in, for example, Meier and Shumlak (2012); Khomenko et al. (2014a).

Here we have assumed that the different charges can be treated as a single fluid (a key assumption in formulating the MHD equations). However, there are situations where treating the different charges (positively charged ions and negatively charged electrons) as different fluids becomes appropriate. The key criteria that need to be satisfied to assume a single, charge-neutral plasma fluid are that the phenomena studied are sufficiently large scale (i.e. larger than the Larmor radius, Debye length and electron mean-free-path) and low frequency (frequency smaller than the plasma and electron gyro-frequency). We have also assumed they are slow (sub-relativistic to allow, for example, the displacement current to be neglected). In this article, we will assume that we are discussing dynamics that satisfy these conditions.

The terms on the RHS of the above equations are the collisional coupling terms. In the momentum equations these are terms that transport momentum between the fluids due to collisions and this is linearly proportional to the velocity drift ($\mathbf{v}_D = \mathbf{v}_n - \mathbf{v}_p$). In the energy equation there is a term that gives both the transfer of kinetic energy and frictional heating of the system, and a term in the temperature difference that forces the fluids to relax to a thermal equilibrium. These momentum and energy transfer terms have a prefactor including the parameter $\alpha_c$ written so they take the form

$$\nu_{in}\rho_p = \alpha_c \rho_n \rho_p = \nu_{ni}\rho_n, \tag{11}$$

with $\nu_{in}$ and $\nu_{ni}$ are the collision frequencies of ions onto neutrals and neutrals onto ions, respectively. Note here there is the assumption that the important collisions in the transfer of momentum and energy between the fluids are those between ions and neutrals and not electrons and neutrals. The form of $\alpha_c$ is approximately given by

$$\alpha_c \approx \frac{1}{m_n + m_i}\frac{8}{3}\sqrt{\frac{2k_B T_r}{\pi m_r}}\left(1 + \frac{9\pi s^2}{64}\right)^{1/2}\Sigma_{in}(\mathbf{v}_D, T_n, T_p), \tag{12}$$

for both hard sphere collisions (Draine, 1986) and charge exchange (Zank et al., 2018), where

$$T_r = \frac{m_i T_n + m_n T_p}{m_n + m_i}, \tag{13}$$

$$m_r = \frac{m_n m_i}{m_n + m_i}, \tag{14}$$

$$s = v_D \left(\frac{2k_B T_r}{m_r}\right)^{-1/2}, \tag{15}$$

$$v_D = |\mathbf{v}_n - \mathbf{v}_p|, \tag{16}$$

with $T_n$ and $T_p$ the neutral and plasma temperatures, $m_n$ and $m_i$ the mass of the neutral and ion particles involved in the collisions and $k_B$ Boltzmann's constant. Though it is common to ignore the term that allows collisions to be driven by velocity drifts, it can become important in highly dynamic systems (e.g. Murtas et al., 2021). The collisional cross-section ($\Sigma_{in}$) is generally larger for charge exchange interactions than hard-sphere collisions but these cross-sections can vary greatly with the collisional velocity (e.g. Vranjes and Krstic, 2013).

In this review article we will focus on the two-fluid approximation and results from these calculation. There have been a range of studies looking at solar dynamics using this approximation. The study of drift velocities in prominences by Gilbert et al. (2002) predicts that there is a slow mass drainage of prominence material (with heavier elements draining faster). This was confirmed numerically by Terradas et al. (2015), who highlighted that the magnitude of the drift velocity is inversely proportional to the magnitude of the collision frequency. The observational study of Gilbert et al. (2007) seemed to confirm that the rate of draining of material from a filament was species dependent. There have also been a number of studies looking at the role of frictional heating by ion neutral drift for waves in the solar chromosphere (Kuźma et al., 2019; Wójcik et al., 2020).

It is important to note that for the two-fluid approximation to hold, it is necessary that the particles in a fluid have significantly more interactions with particles in their own fluid than that of the other. If this is not the case, then assuming that the mean-free-path of a particle is controlled by interactions with particles of the same species is flawed and can lead to spurious results, e.g. possibly this is the cause of the overly large viscosity found for neutral species in the simulations of Braileanu et al. (2021a).

### 2.1. Connecting between the two-fluid and single fluid approximations

A big question that faces anyone trying to model partially ionised plasma dynamics in the solar atmosphere is what level of accuracy does one need to model the system, with the main choice being between using a single fluid or two-fluid approximation. The ultimate decision on which is more appropriate will boil down to a question of timescales. We can imagine that if the timescales associated with the collisional coupling are much shorter than the timescale of the dynamics of the system, then it would be natural to expect that the two fluids are strongly coupled such that they only have a small velocity drift and they move almost as a single fluid. Conversely, if the timescale of the dynamics is much shorter than the coupling timescale then the fluids will be decoupled on those timescales and move almost completely independently. There will also be a regime between these where the dynamic and coupling timescales are similar resulting in intermediate or partial coupling of the fluids. But can this be quantified in any way?.

As can be seen from the two-fluid partially ionised plasma equations, the key physical quantities that control the interaction between the two species are the velocity difference (or drift) and the temperature difference. Therefore it makes sense to develop a simple set of equations to perform a reduced analysis of the temporal evolution of these quantities. This can then be used to show how we expect these difference terms to vary with time, and on what timescales this occurs.

We start by looking at the two velocity equations (one for the neutral fluid and one for the plasma fluid):

$$\rho_n \frac{\partial \mathbf{v}_n}{\partial t} + (\rho_n \mathbf{v}_n \cdot \nabla)\mathbf{v}_n + \nabla P_n = -\alpha_c \rho_n \rho_p \mathbf{v}_D, \quad (17)$$

$$\rho_p \frac{\partial \mathbf{v}_p}{\partial t} + (\rho_p \mathbf{v}_p \cdot \nabla)\mathbf{v}_p + \nabla P_p - \nabla \times (\mathbf{B}) \times \mathbf{B} = \alpha_c \rho_n \rho_p \mathbf{v}_D. \quad (18)$$

Here some important terms are missing from the equation, e.g. gravity, but to add these is relatively trivial and would not change the arguments we are about to present. By subtracting the second of these equations from the first, we can derive an equation for the time-evolution of the velocity drift,

$$\frac{\partial \mathbf{v}_D}{\partial t} + \mathbf{v}_n \cdot \nabla \mathbf{v}_n - \mathbf{v}_p \cdot \nabla \mathbf{v}_p = \frac{1}{\rho_p}\nabla p_p - \frac{1}{\rho_p}\nabla \times (\mathbf{B}) \times \mathbf{B}$$
$$- \frac{1}{\rho_n}\nabla p_n - \alpha_c(\mathbf{v}_D, T_n, T_p)(\rho_n + \rho_p)\mathbf{v}_D. \quad (19)$$

This equation shows there are two ways in which a velocity difference can occur at a given point: 1) The force terms (first three terms on the RHS) can create a velocity difference through driving different flows in the different fluids and 2) the transport of flow by the advection terms (on the LHS) can transport different velocity components from different parts of the flow creating a new velocity difference locally. We then have the coupling term, which through simple inspection we can see will result in the exponential damping of any velocity difference.

When looking at this equation, we can see that there are three fundamental timescales (two of which will be discussed in more detail in Section 2.2) of this system, the timescale through which velocity differences are created, the timescale through which they are damped and the timescale through which the collisional coupling terms evolve (i.e. the timescale over which the collision frequencies change). As we are most interested in the consequence of the damping of the velocity drift, for the moment we will focus on the evolution of the velocity drift by investigating its evolution over a period of time that is sufficiently short to "freeze" the dynamic terms and the collision frequencies as approximately constant. Over this short period, the equation we are analysing becomes

$$\frac{d\mathbf{v}_D}{dt} = \mathbf{C}_1 - \alpha_c(\mathbf{v}_D, T_n, T_p)(\rho_n + \rho_p)\mathbf{v}_D, \quad (20)$$

with

$$\mathbf{C}_1 = -\mathbf{v}_n \cdot \nabla \mathbf{v}_n + \mathbf{v}_p \cdot \nabla \mathbf{v}_p + \frac{1}{\rho_p}\nabla p_p - \frac{1}{\rho_p}\nabla \times (\mathbf{B})$$
$$\times \mathbf{B} - \frac{1}{\rho_n}\nabla p_n. \quad (21)$$

Note that $\mathbf{C}_1$ is not treated as a function of time as these terms are assumed to be not changing over the time period we are considering. From this we can analyse the local evolution, over a short time period, of the velocity drift though a linear ODE (which is mathematically considerably simpler than the Eq. 19). The philosophy behind this set of assumptions comes from numerical simulations, where time integrals are performed over a $\Delta t$ where the quantities associated with the fluxes are treated as being fixed over that short time step (determined by the CFL condition). This then allows the equations to be integrated as difference equations. Here we treat it as though only the drift terms can evolve with time, fixing the other terms.

As we are taking both $\alpha_c$ and $\mathbf{C}_1$ as being constant, this allows us to perform a separation of variables in Eq. 20 which leads us to the solution

$$\mathbf{v}_D(t) = \frac{\mathbf{C}_1}{\alpha_c(\rho_n + \rho_p)} + \left(\mathbf{v}_D(0) - \frac{\mathbf{C}_1}{\alpha_c(\rho_n + \rho_p)}\right)\exp(-\alpha_c(\rho_n + \rho_p)t), \quad (22)$$

where $\mathbf{v}_D(0)$ is the velocity drift at $t = 0$. This solution is of the exponential decay of the drift velocity to a constant value. Taking the limit of the applicability of the approximation to be at $t = t'$, we can see that if $\alpha_c(\rho_n + \rho_p)t' \gg 1$ then the velocity difference at $t'$ ($\mathbf{v}'_D$) is given by

$$\mathbf{v}'_D = \frac{\mathbf{C}_1}{\alpha_c(\rho_n + \rho_p)}, \quad (23)$$

i.e. a constant. This is known as the strong coupling limit of the velocity drift, and is applied (often in simplified forms, e.g. Braginskii, 1965; Khomenko, 2020) as a correction to the induction equation giving

$$\frac{\partial \mathbf{B}}{\partial t} = \nabla \times (\mathbf{v}_{CM} \times \mathbf{B} - \xi_n \mathbf{v}'_D \times \mathbf{B}), \quad (24)$$

where $\mathbf{v}_{CM}$ is the centre-of-mass velocity of the two fluids (e.g. Khomenko, 2020). This allows the mass, momentum and energy equations to be summed and solved as a single fluid approximation (e.g. Khomenko et al., 2014b; Nóbrega-Siverio et al., 2020a). A further consequence of this results is that the existence of drift velocities for large (but not infinite) collisional frequencies means that there will be non-zero frictional heating. However, the magnitude of the heating will tend to zero as the coupling gets increasingly stronger (e.g. Hillier, 2019). Conversely, for the case where $\alpha_c(\rho_n + \rho_p)t' \ll 1$ then

$$\mathbf{v}'_D \approx \mathbf{v}_D(0) - \alpha_c(\rho_n + \rho_p)t'\left(\mathbf{v}_D(0) - \frac{\mathbf{C}_1}{\alpha_c(\rho_n + \rho_p)}\right), \quad (25)$$

which will be dominated by the initial velocity drift. This implies that over these timescales the fluids are effectively decoupled.

Now that we have the easy part out the way, we can approach the trickier step of analysing the thermal coupling. Here we start with the two temperature equations for the two fluids

$$\frac{\partial T_n}{\partial t} + \mathbf{v}_n \cdot \nabla(T_n) + (\gamma - 1)T_n\nabla \cdot \mathbf{v}_n = \alpha_c\rho_p \frac{(\gamma-1)\mu_n}{\gamma}\frac{1}{2}\mathbf{v}_D(t)^2 - \alpha_c\rho_p\mu_n T_D \quad (26)$$

$$\frac{\partial T_p}{\partial t} + \mathbf{v}_p \cdot \nabla(T_p) + (\gamma - 1)T_p\nabla \cdot \mathbf{v}_p = \alpha_c\rho_n \frac{(\gamma-1)\mu_p}{\gamma}\frac{1}{2}\mathbf{v}_D(t)^2 + \alpha_c\rho_n\mu_p T_D, \quad (27)$$

with $T_D = T_n - T_p$. This allows us to formulate an equation for the temperature difference

$$\frac{dT_D}{dt} = C_3 + \alpha_c(\rho_p\mu_n - \rho_n\mu_p)\frac{(\gamma-1)}{\gamma}\frac{1}{2}\mathbf{v}_D(t)^2 - \alpha_c(\rho_p\mu_n + \rho_n\mu_p)T_D, \quad (28)$$

with $C_3 = -\mathbf{v}_n \cdot \nabla(T_n) - (\gamma - 1)T_n\nabla \cdot \mathbf{v}_n + \mathbf{v}_p \cdot \nabla(T_p) + (\gamma-1)T_p\nabla \cdot \mathbf{v}_p$. Again we will be analysing the system over a sufficiently short time period such that the terms in $C_3$ and the collision frequencies can be treated as approximately constant. One thing that can be pointed out here is that $C_3$ could incorporate any heating term (e.g. Ohmic heating, viscous heating or even radiative losses) which can be considered approximately constant over the timescale of interest.

The solution to this equation will have the form $T_D = p(t) + C\exp(-\alpha_c(\rho_p\mu_n + \rho_n\mu_p)t)$ where $p(t)$ is the particular integral and $C$ is a constant we can determine once $p(t)$ has been found. After some algebra this gives

$$T_D(t) = T_D(0) - D\exp(-\alpha_c(\rho_p\mu_n + \rho_n\mu_p)t) \quad (29)$$

$$+ \left(C_3 + \frac{\mathbf{C}_1^2(\rho_p\mu_n - \rho_n\mu_p)}{\alpha_c(\rho_n+\rho_p)^2}\frac{\gamma-1}{2\gamma}\right)(\alpha_c(\rho_p\mu_n + \rho_n\mu_p))^{-1} \quad (30)$$

$$+ A\exp(-\alpha_c(\rho_n + \rho_p)t) + B\exp(-2\alpha_c(\rho_n + \rho_p)t)$$

$$A = \frac{\rho_p\mu_n - \rho_n\mu_p}{(\rho_p\mu_n + \rho_n\mu_p - \rho_n - \rho_p)}\frac{\gamma-1}{2\gamma}\frac{\mathbf{C}_1}{\alpha_c(\rho_n+\rho_p)} \times \left(\mathbf{v}_D(0) - \frac{\mathbf{C}_1}{\alpha_c(\rho_n+\rho_p)}\right) \quad (31)$$

$$B = \frac{\rho_p\mu_n - \rho_n\mu_p}{(\rho_p\mu_n + \rho_n\mu_p - 2\rho_n - 2\rho_p)}\frac{\gamma-1}{2\gamma}\left(\mathbf{v}_D(0) - \frac{\mathbf{C}_1}{\alpha_c(\rho_n+\rho_p)}\right)^2 \quad (32)$$

$$D = A + B + \left(C_3 + \frac{\mathbf{C}_1^2(\rho_p\mu_n - \rho_n\mu_p)}{\alpha_c(\rho_n+\rho_p)^2}\frac{\gamma-1}{2\gamma}\right)(\alpha_c(\rho_p\mu_n + \rho_n\mu_p))^{-1}. \quad (33)$$

There is one very interesting and often overlooked consequence here, in the limit of strong coupling the temperature difference does not reduce to zero (a standard assumption applied when deriving a single fluid model of partially ionised plasma dynamics, Braginskii, 1965). In fact, it becomes:

$$T_D(t) = \left(C_3 + \frac{\mathbf{C}_1^2(\rho_p\mu_n - \rho_n\mu_p)}{\alpha_c(\rho_n+\rho_p)^2}\frac{\gamma-1}{2\gamma}\right)(\alpha_c(\rho_p\mu_n + \rho_n\mu_p))^{-1}. \quad (34)$$

Here we can see that we can have a temperature difference that, like the velocity drift, is inversely proportional to the rate of the collisional coupling between the fluids. The temperature difference in any region is driven by advection of fluids of different temperatures into a region, different levels of compression of the fluids and the frictional heating created by the drift between the two species. Though it is likely that in many instances these effects are not so large, for example highly compressible dynamics like shocks are able to drive large temperature differences in two-fluid models but this is completely ignored in single fluid approximations of this phenomenon. This formulation may allow a more realistic temperature variance in single fluid calculations of the solar atmosphere where the correctly modelling the electron temperature is important to calculate collisional ionisation and recombination rates.

The alternate situation of very weak coupling is also meaningful to understand. Similarly to the velocity difference, the temperature difference is then determined by its initial value, i.e. $T_D(t) \approx T_D(0)$ with a small correction that is linear in time. In this case the temperature difference does not vary greatly over the dynamic timescale, which implies that for fast changes in temperature the fluids are thermally decoupled.

An interesting point of note is this analytic formulation for $\mathbf{v}_D$ and $T_D$ can be used directly in the numerical simulations of two-fluid phenomena by replacing the difference terms in the stiff terms on the RHS of the two-fluid equations presented in the two-fluid equations. This allows these terms to be directly integrated with time over a given timestep analytically due to their exponential form. Therefore, instead of considering using implicit methods, this form of exponential integrator may also prove to be a versatile tool for numerically modelling partially ionised plasma phenomena in a complex, partially ionised atmosphere. A simplified version of this form has been implemented in the (PIP) code (Hillier et al., 2016), but more work needs to be done to see if the it can be used as part of efficient, accurate two-fluid simulations of the Chromosphere.

## 2.2. The multiscale nature of partially ionised plasma dynamics

The previous subsection has shown that the ratio of the dynamic timescale to the collision timescale changes fundamentally how a system responds to any motion, with a key point of comparison being ratio of the characteristic timescales of the collisions to the dynamic timescale. This then leads us to the following question: What is the dynamic timescale of MHD phenomenon in the solar atmosphere?.

An analogy can be drawn between this ratio and the Deborah number, often used in rheology, which is given by the ratio of the relaxation time of a non-Newtonian fluid (which connects with our collisional time) to the timescale of the experiment (which connects to our dynamic timescale). The name for the Deborah number comes from the old testament, with the line from the song by the prophetess Deborah: "The mountains flowed before the Lord". This name was used in the relaxation of fluids as given a sufficiently long relaxation time (e.g. geological timescales) even mountains can be observed to flow like a fluid. When performing experiments on non-Newtonian fluids it is important to compare data for experiments with similar Deborah numbers as the dynamics fundamentally changes as this number is varied. With Newtonian viscous flow-like behaviour at low Deborah number and non-Newtonian behaviour at high Deborah number.

As there are fundamentally many different timescales associated with dynamics in the solar atmosphere, considering the ratio of the collision frequency to the timescale dynamics of interest is of great importance to decide whether single or two-fluid approximations are most appropriate. However, it is of high importance to note that multiple timescales can all naturally exist in the same system at the same time, meaning the ratio of timescales can be both small and large, and it may also be that to correctly model the scale of greatest interest a wide variety of timescales must be considered. One could say that in solar physics it is important to simultaneously understand both solid and flowing mountains (or how high-frequency, two-fluid dynamics feeds back on larger-scale, slower dynamics). In the rest of this section we will provide some broadstrokes examples of the fundamentally multi-time-scale nature of solar dynamics.

### 2.2.1. Alfvén waves

Due to both their ubiquitous nature in the solar atmosphere, and the relative simplicity in providing analytical formula, it is good to start with waves. Focusing on the Alfvén wave (following Soler et al., 2013b), we can determine the frequency for an Alfvén wave in a two-fluid medium from the solutions of the following cubic equation

$$\omega^2(i\alpha_c(\rho_n + \rho_p) + \omega) - V_A^2 k^2 \cos^2\theta(\omega + i\alpha_c\rho_p) = 0, \qquad (35)$$

where $V_A$ is the Alfvén speed in the plasma, e.g. $V_A = B/\sqrt{\rho_p}$. As this is a cubic equation its solutions give the frequencies for three different waves. These are the forward and backward propagating Alfvén waves, and a neutral shear wave. Due to the non-propagating nature of the neutral shear wave (and arguments on symmetry) we would not expect this extra wave to have any real frequency and analysis shows that this is purely damped (Soler et al., 2013b). For the two other solutions, generally these are the forward and backward propagating, damped Alfvén waves though there are situations when these both become purely damped.

It is rather simple to look at the limits for Eq. 35. If $\alpha_c(\rho_n + \rho_p)$ is small compared to $\omega$ then we just have

$$\omega(\omega^2 - V_A^2 k^2 \cos^2\theta) \approx 0, \qquad (36)$$

i.e. one zero frequency neutral shear wave and two Alfvén waves, with the Alfvén wave frequency determined by the plasma Alfvén speed. For the photosphere, with its very high collision frequency this regime would occur when $Re(\omega)$ is at least greater than $10^9\,s^{-1}$, though nearer the top of the chromosphere this becomes $Re(\omega) \gtrsim 10^3\,s^{-1}$ due to the drastically smaller collision frequency at this height. For the opposite extreme, $\alpha_c \rho_p$ is large, Eq. 35 reduces to

$$\omega^2 - V_A^2 k^2 \frac{\rho_p}{\rho_n + \rho_p} = 0, \quad (37)$$

here we have two waves, forward and backward propagating Alfvén waves where the wave frequency is determined by the bulk Alfvén speed, i.e. $V_{At} = B/\sqrt{\rho_n + \rho_p}$. Therefore, two-fluid Alfvén waves can behave like a single-fluid MHD Alfvén wave or like an Alfvén wave where the plasma doesn't know there is a neutral fluid (or somewhere in between). Either of these extremes can be easily excited. As a note, there has also been some detailed investigations into compressional MHD waves, and the reader is directed to Soler et al. (2013a) and Alharbi et al. (2022) for more details on these cases. Here we will look in greater detail at shock waves as a fundamentally multi-scale phenomenon (see Section 3).

*2.2.2. Instabilities and magnetic reconnection*

Moving on from waves it is natural to consider instabilities. When looking at the linear stability, similar arguments can be put forward as those for waves: if the growth rate is large enough then the two fluids will decouple during the growth of the instability, but for small growthrates they will behave like a single fluid. However, it can be difficult, sometimes impossible, to know a priori which of these regimes will be the most important for understanding the dynamics of interest. Nor is it clear a priori whether two-fluid effects will manifest themselves in an important contribution to the non-linear dynamics of the instability. We will look at instabilities in greater detail in Section 4.

Another important dynamic phenomenon of the solar atmosphere is magnetic reconnection, with observations showing that fast, bursty reconnection can occur in the solar chromosphere (Singh et al., 2011). The fundamental timescale of reconnection is the Alfvén time based on the half-length of the current sheet as it gives a timescale for material to be ejected from the reconnection region. However, the dynamics in a reconnecting current sheet are more complex with plasmoids forming through resistive instabilities (Zweibel, 1989) and then interacting to drive multi-scale, potentially fractal reconnection (Singh et al., 2015; Singh et al., 2019). This naturally results in a multi-scale process that can scale the dynamic frequencies covered by single-fluid, two-fluid and into kinetic dynamics (Singh et al., 2015).

*2.2.3. Turbulence*

Finally, it is very interesting to look at turbulence. This is the archetypal example of multi-time-scale, multi-spatial-scale dynamics in a fluid system, and both MHD and partially ionised plasma turbulence also have this property. As such the existence of a turbulence cascade can take energy from large-scale, well-coupled flows and inject it in scales that are becoming more and more decoupled (e.g. Burkhart et al., 2015). This is easy to see from a simple argument, the frequency of motions at any scale in a steady-state turbulent cascade can be approximated by

$$f_{\rm turb} = \left(\frac{\epsilon}{L^2}\right)^{1/3}, \quad (38)$$

where $\epsilon$ is the energy cascade rate of the turbulent cascade and $L$ is the dynamic scale under consideration. This frequency increases as the scale under consideration of the turbulent cascade decreases. Assuming that viscosity does not truncate the cascade too soon, there will exist scales $L_{\nu_{ni}}$ and $L_{\nu_{in}}$ such that the turbulent frequency surpasses the neutral ion and ion neutral collision frequencies, respectively. Therefore, during a turbulent cascade in partially ionised plasma it would be natural for highly-coupled dynamics to become decoupled requiring two-fluid modelling.

In the following sections we will look in more detail at a few examples where the natural existence of multiple scales of interest or multi-scale dynamics make them key areas in which systems can naturally exhibit two-fluid partially ionised plasma effects. We will focus on shocks (Section 3) and and linear and non-linear instabilities (Section 4) in two-fluid systems.

**3. The role of partially ionised plasma in shocks**

Shocks are highly-nonlinear, highly-compressible waves characterised by very sharp transitions in density, velocity, pressure and, for MHD shocks, magnetic field around the shock front. This sharp transition in physical quantities results from the magnitude of the flow in a system abruptly changing from above a characteristic wave speed to below, with the three wave-speeds of MHD (slow, fast and Alfvén) leading to three different shock jumps (e.g. Delmont and Keppens, 2011). However, as we have seen for partially ionised plasmas when considering Alfvén waves (see Section 2.2 or Soler et al., 2013b) or for the slow and fast magnetoacoustic wave modes (e.g. Soler et al., 2013a; Alharbi et al., 2022) there are multiple version of the same wave speed. As such a partially ionised plasma system could possess multiple different flavours of the three types of shocks (potentially simultaneously).

The possibility of this multiple flavours of different shocks has to be understood in the context of their lifetime. Once a shock has propagated a sufficient distance, the upstream and downstream states should couple together making a system that on large scales behaves like strongly coupled MHD. The work of Hillier et al. (2016) studied the evolution of a switch-off shock (a type of slow-mode MHD shock) in a two-fluid system highlighting just such an effect.

Initially a strong shock was created in the plasma, but as this propagated the neutrals began to respond, forcing this plasma shock to evolve towards one that obeyed a shock transition of a highly coupled system. This leads to fundamentally two timescales in the shock problem, the high-frequency timescales of the shock front, and the evolving timescales relating to the coupling between the fluids in the regions around the shock, this latter timescale is related to the length of time a shock has propagated. This makes shocks a good example where two-fluid dynamics may connect to the larger-scale, slower-evolving dynamic evolution of the solar atmosphere.

Due to the importance shocks are believed to play in the heating of the lower solar atmosphere (e.g. Hollweg et al., 1982; Brady and Arber, 2016) and their clear observations (e.g. Chae et al., 2018; Houston et al., 2018; Houston et al., 2020) it is important to understand the nature of shocks in the partially ionised layers of the solar atmosphere. The study of partially ionised plasma shocks in relation to the solar atmosphere is a relatively new field, with a lot of the understanding coming as an extension of the work looking at shocks in the interstellar medium. For a review on this topic see, for example, Draine and McKee (1993). In this section we review some of the studies of particular importance for understanding two-fluid effects in shocks in the lower solar atmosphere.

### 3.1. The PIP shock jump

One of a theorists fundamental tools to analyse and understand a given shock are the shock jump conditions (e.g., Goedbloed et al., 2010). These are a set of 1D conditions on the mass, momentum, energy and magnetic fluxes to enforce conservation of these quantities across a shock front. The ideal MHD equations can be moved to the Hoffman-Teller frame (the shock frame with zero electric field both upstream and downstream of the shock) then integrated to yield:

$$[\rho v_\perp]^u_d = 0, \quad (39)$$

$$\left[\rho v_\perp^2 + P + \frac{B_\parallel^2}{2}\right]^u_d = 0, \quad (40)$$

$$[\rho v_\perp v_\parallel - B_\perp B_\parallel]^u_d = 0, \quad (41)$$

$$\left[v_\perp \left(\frac{\gamma}{\gamma-1}P + \frac{1}{2}\rho v^2\right)\right]^u_d = 0, \quad (42)$$

$$[B_\perp]^u_d = 0, \quad (43)$$

$$[v_\perp B_\parallel - v_\parallel B_\perp]^u_d = 0, \quad (44)$$

where this notation means that

$$[Q]^u_d \equiv Q^u - Q^d \quad (45)$$

for any quantity $Q$. These can be reduced to a single equation that gives the possible stable shock jumps for a specified upstream plasma beta and upstream angle of magnetic field in terms of the upstream and downstream Alfvén Mach numbers (Hau and Sonnerup, 1989):

$$A_x^{u2} = \left[A_x^{d2}\left(\frac{\gamma-1}{\gamma}\left(\frac{\gamma+1}{\gamma-1} - \tan^2\theta\right)(A_x^{d2}-1)^2\right.\right.$$
$$\left.\left. + \tan^2\theta\left(\frac{\gamma-1}{\gamma}A_x^{d2} - 1\right)(A_x^{d2}-2)\right) - \frac{\beta}{\cos^2\theta}(A_x^{d2}-1)^2\right]$$
$$/\left[\frac{\gamma-1}{\gamma}\frac{(A_x^{d2}-1)^2}{\cos^2\theta} - A_x^{d2}\tan^2\theta\left(\frac{\gamma-1}{\gamma}A_x^{d2}-1\right)\right], \quad (46)$$

for the Alfvén Mach number $A_x = v_x \frac{\sqrt{\rho}}{B_x}$ evaluated in the upstream and downstream media. This is a very powerful tool as it allows us to calculate the types of shocks that can exist in a system from minimal details of the upstream medium. Note that this form is equivalent to the shock adiabatic including a linear (trivial) solution of $A_x^u = A_x^d$, i.e., no shock.

This analysis can be extended to the two-fluid system given by Eqs. 1,4,5,10 (Snow and Hillier, 2019). For simplicity, the ion and neutral species are assumed to be coupled by thermal collisions only (with gravity, viscosity, magnetic diffusion etc. neglected). In the Hoffman-Teller frame, the two-fluid PIP equations become:

$$\rho_n v_{\perp n} = \text{const}, \quad (47)$$

$$\rho_n v_{\perp n} v_{\parallel n} = -I_1 + \text{const}, \quad (48)$$

$$\rho_n v_{\perp n} v_{\perp n} + P_n = -I_2 + \text{const}, \quad (49)$$

$$v_{\perp n}\left(\frac{\gamma}{\gamma-1}P_n + \frac{1}{2}\rho_n v_n^2\right) = -I_3 + \text{const}, \quad (50)$$

$$\rho_p v_{\perp p} = \text{const}, \quad (51)$$

$$\rho_p v_{\perp p} v_{\parallel p} - B_\perp B_\parallel = I_1 + \text{const}, \quad (52)$$

$$\rho_p v_{\perp p} v_{\perp p} + P_p + \frac{B^2}{2} = I_2 + \text{const}, \quad (53)$$

$$v_{\perp p}\left(\frac{\gamma}{\gamma-1}P_p + \frac{1}{2}\rho_p v_p^2\right) = I_3 + \text{const}, \quad (54)$$

$$v_{\perp p} B_\parallel - v_{\parallel p} B_\perp = 0, \quad (55)$$

$$B_\perp = \text{const}, \quad (56)$$

$$I_1 = \int \alpha_c(T_n, T_p)\rho_n \rho_p (v_{\parallel n} - v_{\parallel p}) \mathrm{d}\perp, \quad (57)$$

$$I_2 = \int \alpha_c(T_n, T_p)\rho_n \rho_p (v_{\perp n} - v_{\perp p}) \mathrm{d}\perp, \quad (58)$$

$$I_3 = \int \alpha_c(T_n, T_p)\rho_n \rho_p$$
$$\left[\frac{1}{2}(v_n^2 - v_p^2) + \frac{1}{\gamma-1}R_g(T_n - T_p)\right]\mathrm{d}\perp. \quad (59)$$

Note that in these equations, integral terms $(I_1, I_2, I_3)$ exist that govern the exchange of momentum and energy between the ion and neutral species. However, these integral terms can not be evaluated easily. Instead, due to the conservative nature of the system, the integral terms can be removed by adding the neutral and ion equations together, reducing the equations to:

$$\rho_n v_{\perp n} + \rho_p v_{\perp p} = \text{const}, \tag{60}$$

$$\rho_n v_{\perp n} v_{\|n} + \rho_p v_{\perp p} v_{\|p} - B_\perp B_\| = \text{const}, \tag{61}$$

$$\rho_n v_{\perp n} v_{\perp n} + P_n + \rho_p v_{\perp p} v_{\perp p} + P_p + \frac{B^2}{2} = \text{const}, \tag{62}$$

$$v_{\perp n}\left(\frac{\gamma}{\gamma-1}P_n + \frac{1}{2}\rho_n v_n^2\right) + v_{\perp p}\left(\frac{\gamma}{\gamma-1}P_p + \frac{1}{2}\rho_p v_p^2\right)$$
$$= \text{const}, \tag{63}$$

$$v_{\perp p} B_\| - v_{\|p} B_\perp = 0, \tag{64}$$

$$B_\perp = \text{const}. \tag{65}$$

The equations can be further simplified by expressing the partial pressures and densities as a total value using the neutral fraction $\xi_n$:

$$\rho_n = \xi_n \rho_t, \tag{66}$$

$$\rho_p = (1-\xi_n)\rho_t, \tag{67}$$

$$P_n = \frac{\xi_n}{\xi_n + 2(1-\xi_n)}P_t, \tag{68}$$

$$P_p = \frac{2(1-\xi_n)}{\xi_n + 2(1-\xi_n)}P_t. \tag{69}$$

Substituting these into Eqs. 60,65 and furthermore imposing an additional constraint such that upstream and downstream of the shock the drift velocity equals zero ($v_{\|p} = v_{\|n} = v_\|$ and $v_{\perp p} = v_{\perp n} = v_\perp$):

$$\rho_t v_\perp = \text{const}, \tag{70}$$

$$\rho_t v_\perp v_\| - B_\perp B_\| = \text{const}, \tag{71}$$

$$\rho_t v_\perp v_\perp + P_t + \frac{B^2}{2} = \text{const}, \tag{72}$$

$$v_\perp\left(\frac{\gamma}{\gamma-1}P_t + \frac{1}{2}\rho_t v^2\right) = \text{const}, \tag{73}$$

$$v_\perp B_\| - v_\| B_\perp = 0, \tag{74}$$

$$B_\perp = \text{const}. \tag{75}$$

Eqs. 70,75 are identical to the MHD jump equations (Eqs. 39,44) except here the density and pressure are given by their total quantities $\rho_t, P_t$. Therefore, the solution to these equations is identical to the Hau and Sonnerup, 1989 solution for MHD (Eq. 46), independent of the neutral fraction. It should be noted that this is only true over the larger shock structure and inside the shock there is substructure that is highly dependent on the collisional effects. This result should not be of any surprise. The collisional coupling terms conserve both momentum and energy, so any model that looks at the overall conservation of these quantities over both fluids will reduce to the MHD solution sufficiently upstream and downstream of the shock.

### 3.2. Finite Width shocks and their substructure

In the previous subsection we have seen that at least for a steady-state shock front the inclusion of partially ionised plasma effects does not change the overall shock jump as this is determined by conservation of total mass, momentum and energy as well as magnetic flux. However, the inclusion of dissipative terms means that the discontinuous shock front that exists in ideal, inviscid MHD and inviscid hydrodynamics can no longer be sustained. Instead, the shock now has a finite width that is determined by a physical length scale of the dissipation terms (e.g. Hau and Sonnerup, 1989; Draine and McKee, 1993).

As the ionised and neutral species enter the shock they decouple before recoupling as they pass through the shock, see Fig. 2. This allows large drift velocities to develop in the shock, which results in frictional heating within the finite shock width. The thickness of the shock naturally scales with linearly with the inverse of the magnitude of the collisional coupling (e.g. Draine and McKee, 1993; Snow and Hillier, 2021b). However, for other parameters of the system, the correlation to the shock width is not so straight forward. For example, the scaling of the thickness of the two-fluid, switch-off slow-mode shock to the ionisation fraction ($\xi_i$) was found to be proportional to $\xi_i^{-1.2}$ (Hillier et al., 2016).

What happens inside this finite width, as the fluids decouple and recouple, can be very complex, though the basic steady-state solutions can be generally categorised as either continuous (C-) shocks or jump (J-) shocks (Draine and McKee, 1993). For the C-shocks, all the physical quantities vary smoothly over the shock jump. However for the J-shocks there exists a further shock jump inside the larger shock transition. Often this is a hydrodynamic shock in the neutral fluid (e.g. Draine and McKee, 1993; Hillier et al., 2016), which can form as a result of the supersonic velocity forming in the neutrals through being dragged by the plasma (Draine and McKee, 1993) or pushing ahead (Hillier et al., 2016) depending on the type of shock. Transient, though long lived, shocks can also form as the system evolves towards its steady state, with intermediate shocks found to occur in switch-off shocks (Snow and Hillier, 2019). These were found to be created as the plasma gets accelerated in the shock becoming super-Alfvénic, allowing and intermediate shock transition to occur, accompanied by the tell-tale reversal of the magnetic field. This reversal of the magnetic field cannot be supported by the neutral pressure resulting in the slow decay of the shock, hence why this feature is transient.

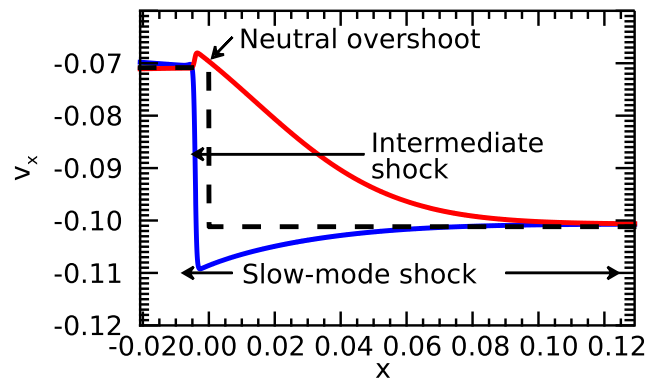

Fig. 2. Finite width of two-fluid switch-off shock showing the plasma (blue) and neutral (red) species from a two-fluid simulation, and the corresponding single-fluid MHD simulation (black), adapted from Snow and Hillier (2019). Note that the upstream and downstream values pair up well with the. MHD simulation. Credit: Snow, B. and Hillier, A. A&A, 626, A46. (2019) reproduced with permission © ESO.

There is an interesting analogy that can be drawn here between the finite width and substructure of partially ionised plasma shocks and similar physics that can be found in hydrodynamic shocks with thermal conduction. In a hydrodynamic shock with thermal conduction, the shock front becomes broad. However, for sufficiently strong shocks at low Prandtl number (ratio of viscosity to thermal conduction), the system develops an internal, isothermal shock (Landau and Lifshitz, 1987; Guidoni and Longcope, 2010). For this shock, the thickness of the global shock transition is fixed by thermal conduction, but the thickness of the internal sub-shock is fixed by viscosity. This is a very similar situation with partially ionised plasma shocks with an internal hydrodynamic shock. the global structure is determined by the coupling length of neutrals to the plasma (and the magnetic field), but the thickness of the internal shock is determined by the shorter lengthscale of coupling of the plasma to the neutrals.

### 3.3. Stratification and partially ionsied plasma shocks

Compressible waves efficiently steepen into shocks as they propagate upwards in the solar atmosphere due to the decreasing density with height creating increasing non-linearity in the wave (e.g., Suematsu et al., 1982). These shocks can form in the lower solar atmosphere, where partial-ionisation plays a key role (for example, umbral flashes Beckers and Tallant, 1969). This leads to the obvious question: what influence does partial ionisation play on shock formation and evolution as a shock propagates upwards through the solar atmosphere?.

When stratification is included, an additional complexity arises. In the two-fluid description described in Section 2, the electrons contribute a pressure (that is equal to the ion pressure) but no density. As such the pressure scale height of the plasma (ions + electrons) is defined as $\Lambda_p = \frac{2T_p}{\gamma g}$, which differs from the neutral pressure scale height $\Lambda_n = \frac{T_n}{\gamma g}$ at equal temperatures. In the two-fluid description described in Section NUMBER, both the ions and the electrons contribute to the plasma pressure scale height, as such it differs from the neutral pressure scale height by a factor of 2, assuming equal temperatures. The different pressure scale heights of ionised and neutral species helps lead to a change in composition of the medium from the mostly neutral photosphere, to the almost entirely ionised transition region. Coupled with this is the reduction in density resulting in weaker coupling between the fluids as the shock propagates upwards. As such, a propagating shock is subject to different levels of coupling with height with different levels of importance for each fluid, which can greatly affect the shock heating of the medium (Niedziela et al., 2021; Zhang et al., 2021). The large neutral fraction in the photosphere results in a relatively strongly coupled media, however as waves or shock propagate further upwards, the coupling becomes weaker (Braileanu et al., 2019a; Braileanu et al., 2019b). This can lead to wave damping (and frictional heating) due to the ion-neutral interactions for both partially ionised waves and shocks. Generally it was found that both smaller period (which excite more two-fluid effects) and larger amplitude (which make shocks form earlier) wave drivers had the strongest damping (Braileanu et al., 2019b). It has been suggested that damping of upward propagating partially-ionised waves alone is a potential way to balance radiative losses from the lower solar atmosphere (Wójcik et al., 2020).

Both waves and shocks can also undergo mode conversion, where one wave mode becomes another, as they propagate upwards in the solar atmosphere (e.g., Schunker and Cally, 2006; Khomenko and Cally, 2019). Since the mode conversion height (where the Alfvén speed equals the sound speed, i.e. $V_A = C_s$) occurs in lower solar atmosphere, it becomes of interest to consider how this process occurs for partially ionised plasmas. In a partially ionised plasma, the height at which mode conversion for waves and shocks occurs can be multiply defined depending on the level of coupling (Snow and Hillier, 2020). For a fully coupled system, the characteristic MHD wave speeds depend on the bulk parameters, i.e., $V_{At} = B/\sqrt{\rho_p + \rho_n}$, $C_{st} = \sqrt{\gamma(P_p + P_n)/(\rho_p + \rho_n)}$, whereas for a completely decoupled systems the MHD wave speeds depend on the isolated plasma properties: $V_A = B/\sqrt{\rho_p}, C_s = \sqrt{\gamma P_p/\rho_p}$ (as discussed in Section 2.2). This provides upper and lower limits limits for where mode-conversion can occurs with the exact height coming from the effective wave speeds calculated by solving a dispersion relation including the coupling frequencies (Soler et al., 2013a).

For shocks in particular, as they pass through the $C_s/V_A = 1$ height, they split into slow- and fast-magnetoacoustic modes. These can either be shocks or smoothed waves depending on the inclination angle of the magnetic field (Pennicott and Cally, 2019). For a partially ionised plasma, this process can be rather non-trivial as multiple waves and shocks can form depending on the level of coupling of the system (Snow and Hillier, 2020). Since the slow-magnetoacoustic plasma speed and the sonic plasma speed are relatively close, the slow component couples fairly readily to the neutral species for moderate levels of coupling. However, the relatively large difference between plasma fast-magnetoacoustic and neutral sonic can result in stark differences to the MHD model. Since the plasma fast magnetoacoustic mode speed is much larger than the neutral sound speed, the neutrals are dragged along by plasma, which can result in localised frictional heating, as well as increasing the finite width of fast-mode shocks. In the study by Snow and Hillier (2020) the finite width of the fast-mode shock was found to exceed the pressure scale height of the system for a wide range of coupling coefficients. Correspondingly, one may expect that fast-mode shocks in the lower solar atmosphere have a finite width of $\approx 300$ km. Note that despite the large shock width, a fast-mode shock transition in still satisfied at the location of the maximum velocity gradient.

*3.4. Stability of Shock Fronts in partially ionised plasma*

In hydrodynamics, shock fronts are considered to be very stable structures, at least in the absence of strong radiation (Laming and Grun, 2002; Grun et al., 1991). This can be understood by the looking at the baroclinic term in the hydrodynamic vorticity equation. As the pressure and density gradients of a shock both point in the same direction, perturbations to hydrodynamic shock fronts typically decay with time. Another way of stating this is the steady state shock solution does not allow any vorticity at the shock front (Zhou et al., 2021).

The inclusion of a magnetic field perpendicular to the shock front fundamentally changes this. The steady state shock front can have vorticity (in the form of jumps in the flow parallel to the shock front) but contact discontinuities cannot (Zhou et al., 2021). A consequence of this is MHD shock fronts may be unstable, but instability of contact discontinuities is suppressed. Fast mode shocks are found to be categorically stable to this instability (provided $\gamma < 3$ Gardner and Kruskal, 1964), whereas the slow-mode shocks may develop the corrugation instability (which corrugates the shock front). The conditions for stability of these shock fronts depends on the Alfvén Mach number and the angle of the magnetic field relative to the shock front (Stone and Edelman, 1995; Édel'Man, 1989; Lessen and Deshpande, 1967).

For a parallel shock, where the velocity and magnetic field are aligned, the corrugation instability grows unconditionally in MHD (Stone and Edelman, 1995), however the corresponding HD shock is unconditionally stable. For a partially ionised system, it is less easy to determine if the shock front will be stable or unstable to the corrugation instability. An argument of timescales tells us that both an instability that grows on timescales much shorter than the coupling time and much longer than the coupling time will experience different MHD limits, allowing them both to be unstable. However, a stability range exists whereby the coupling allows the neutral component to suppress the instability (Snow and Hillier, 2021b). Fig. 3 shows the evolution of the corrugation instability in a partially ionised plasma with different levels of coupling, and MHD simulations of the zero and infinite coupling cases. The MHD simulations are unstable, whereas the two-fluid simulations can become stable for finite coupling. A caveat here is that this has only been studied in 2D parallel shocks. Shocks that are stable in 2D can become unstable

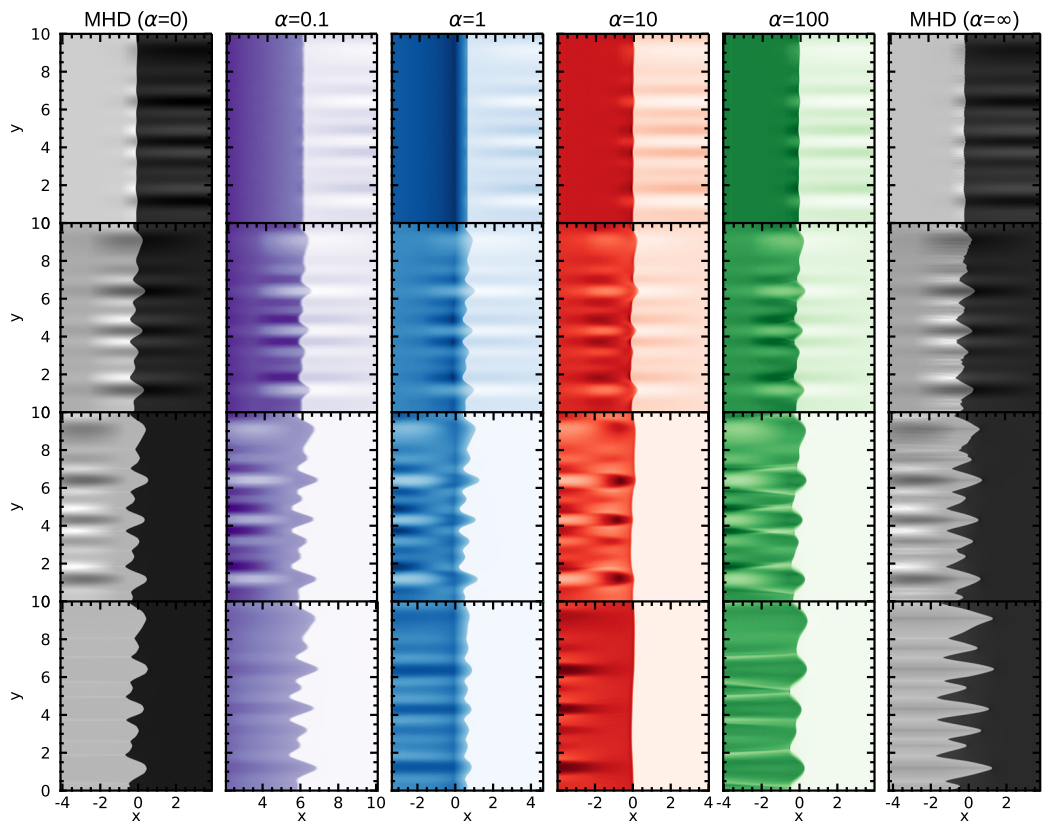

Fig. 3. Time series showing the evolution of the corrugation instability in MHD and partially ionised systems for different levels of coupling, taken from Snow and Hillier (2021b). Credit: Snow, B. & Hillier, A., MNRAS Volume 506, Issue 1, pp.1334-1345, (2021b), Figure 10 reproduced with permssion

to the corrugation instability in 3D due to wide varieties of angles that form between the velocity and magnetic field (Stone and Edelman, 1995; Édel'Man, 1989).

Propagating shocks in the solar atmosphere are almost guaranteed to encounter perturbations and hence corrugate due to the inherent inhomogeneity of the atmosphere. In the lower solar atmosphere, which is known to be both partially ionised and abundant with shocks, one can use characteristic properties of the media to estimate the stability range with respect to perturbation wavelength. Applying this to a sunspot, where the field is reasonably aligned with the direction of propagation and hence the analysis of partially ionsied parallel shocks in Snow and Hillier (2021b) becomes applicable, one can estimate that shocks are stable to perturbations between 0.6 and 56 km.

### 3.5. Shock ionisation and cooling

The rapid changes in density and temperature across a shock lead to locally enhanced ionisation and recombination rates (e.g., Carlsson and Stein, 2002). As such, these effects become critical in modelling partially ionised shocks and omitting ionisation and recombination can lead to a simultaneous over-prediction of shock heating and under-prediction of energy dissipation (Zhang et al., 2021).

In its simplest form, ionisation and recombination appear in the two-fluid equations as an exchange of mass, momentum and energy between the neutral and ionised species such that these quantities are all conserved. Therefore, shock jump equations can be written including these terms and the same analysis can be performed as in Section 3.1 to show that the pre- and post-shock states can be modelled as MHD jumps (Snow and Hillier, 2021a). However, this is only part of the picture. A more realistic model for ionisation and recombination of hydrogen involves radiative losses. Ionisation as a three-body process involves a free electron expending energy to release a bound electron, hence energy is lost from the macroscopic plasma species. The energy lost is proportional to the ionisation rate multiplied by the energy required to release the bound electron (for example, 13.6 eV for ground state hydrogen). Including the ionisation loss modifies the plasma energy equation in the two-fluid model as:

$$\frac{\partial}{\partial t}\left(e_p + \frac{\mathbf{B}^2}{2}\right) + \nabla \cdot [\mathbf{v}_p(e_p + P_p) - (\mathbf{v}_p \times \mathbf{B}) \times \mathbf{B}]$$
$$= \alpha_c \rho_n \rho_p \left[\frac{1}{2}(\mathbf{v}_n^2 - \mathbf{v}_p^2) + \frac{1}{\gamma-1}\left(\frac{P_n}{\rho_n} - \frac{1}{2}\frac{P_p}{\rho_p}\right)\right]$$
$$- \frac{1}{2}\left(\Gamma_{rec}\rho_p \mathbf{v}_p^2 - \Gamma_{ion}\rho_n \mathbf{v}_n^2\right) - \phi_I + A_{heat}$$
$$- \frac{1}{(\gamma-1)}\left(\frac{1}{2}\Gamma_{rec}P_p - \Gamma_{ion}P_n\right),$$
(76)

where $A_{heat}$ is an arbitrary heating term included to balance the system. Note that since ionisation potential loss terms are a energy sink in the plasma only, the neutral energy equation remains unchanged.

A consequence of an energy loss term is that the energy equation is no longer conservative. As such, it cannot be used to construct shock jump relations. Instead, for specific types of cooling, requirements on the heating and cooling terms balancing sufficiently upstream and downstream of the shock can be used to create a semi-analytical description of the stable shock solutions (for example the empirical ionisation and recombination formulae used in Snow and Hillier, 2021a), as shown in Fig. 4. It is obvious that including the ionisation potential loss into the equations has resulted in a very different set of solutions to the MHD (or conservative two-fluid) solutions. This is in contrast to the conservative equations which reduce to MHD sufficiently far from the shock. The general shape of the solutions are both cubic with intersections with the linear (trivial) solution occurring near the slow, Alfvén and fast wave speeds.

As a result of their being cooling the shocks are much more compressible: in MHD the compressiblity limit is given by $r = \rho^d/\rho^u = (\gamma+1)/(\gamma-1)$ whereas here, the compression across the shock can be far greater. Taking for example the solution at $A_x^u = 1$, (i.e., a switch-off slow mode shock), the compression across the shock can be calculated as $r = A^{u2}/A^{d2} \approx 1/0.05 = 20$, which is far greater than the MHD limit. This compression has been confirmed in the simulations of Snow and Hillier (2021a).

As a consequence of this cooling, it has also been found that the compressible shocks in this model will always be cooler downstream of the shock than upstream. This is because the requirements for equilibrium is that the losses are balanced by the background heating term. Equating the heating and the losses, one sees that a compression of the plasma (and the increased density this implies) can only be balanced by a post-shock reduction of temperature.

Until now, the physics we have been discussing would apply to a discontinuous shock as much as a partially ionised plasma shock with a finite width. However, one large difference will exist for these different scenarios: in a shock with a finite width there is the potential for cooling to happen inside the shock front stopping the plasma reaching the maximum predicted temperature of the ideal shock jump. A manifestation of this effect is seen in molecular clouds where molecules that should be dissociated due to the temperature rise predicted by the theoretical MHD shock were found to still exist in the post shock media. The radiative losses inside the shock were found to be crucial in reducing the maximum temperature obtained within the shock and hence allow these molecules to survive the shock (Draine and McKee, 1993). For a hydrogen plasma more relevant for the solar atmosphere, the simulations of Snow and Hillier (2021a) showed that the radiative losses will reduce the maximum temperature obtained within the finite width of the shock. Fig. 5 shows the plasma and neutral temperatures across the finite-width of a switch-off shock for different recombination timescales. Within the shockfront, the ionisation and recombination rates are increased by several orders of magnitude such that even when the background recombination timescale is much longer than the evolution time of the system, the plasma within the shock has cooled to slightly below the MHD limit (orange line). For larger background rates, the cooling is more extreme. This could have important consequences for heating, ionisation and line formation around shocks in the chromosphere.

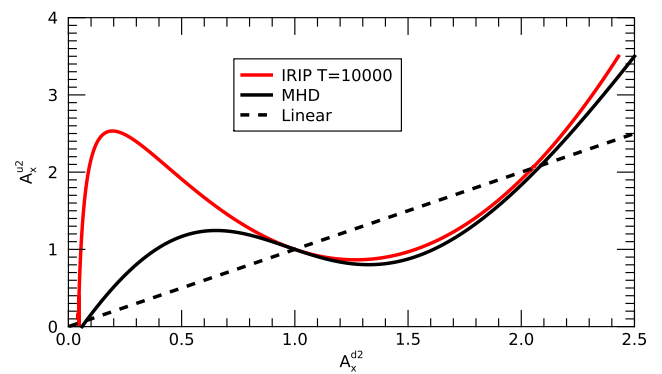

Fig. 4. Shock jump solutions for MHD (black) and two-fluid with ionisation potential losses (red) relating the upstream ($^u$) and downstream ($^d$) Alfvén Mach numbers $A_x$ (taken from Snow and Hillier, 2021a). The trivial solution ($A_x^d = A_x^u$) exists for both sets of equations. The reference upstream parameters are: $T_0 = 10000$ K, $\beta = 0.1$, $\theta = \pi/4$, and $\gamma = 5/3$. Credit: Snow, B., & Hillier, A., A&A, 645, A81 (2021a), reproduced with permission © ESO.

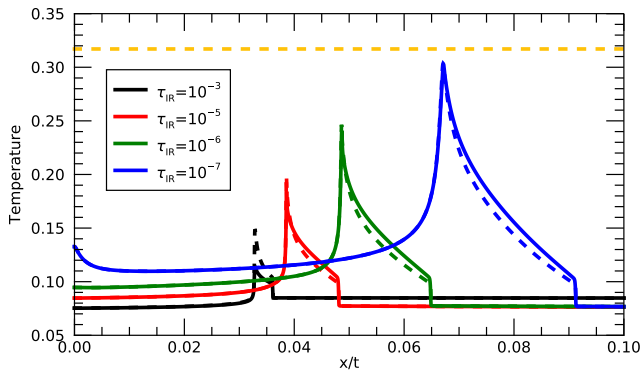

Fig. 5. Plasma (solid) and neutral (dashed) temperatures for different background recombination rates after 10000 time units (taken from Snow and Hillier, 2021a). A background value of $\tau_{IR} = 10^{-7}$ corresponds to a background recombination time scale of $10^7$ time units, however rates are significantly enhanced within the shock. The orange dashed line shows the downstream MHD temperature. Credit: Snow, B., & Hillier, A., A&A, 645, A81 (2021a), reproduced with permission © ESO.

## 4. Instabilities in solar partially ionised plasma

Instabilities are another physical process where we can find the importance of partially ionised plasma effects. This might be through changes in the linear growth rate (e.g. Díaz et al., 2012; Soler et al., 2012) or important nonlinear dynamics driving large velocity drifts (e.g. Braileanu et al., 2021b). In this section we review some of the key results that have been found to date for instabilities in two-fluid modelling relating to the solar atmosphere. For details on a wider range of instabilities and their applications to different astrophysical systems the readers are referred to Soler and Ballester (2022).

### 4.1. Linear instabilities in a partially ionised plasma

A brief outline of the role of partial ionisation in linear stability theory was provided in Section 2.2. As with linear wave theory there is a simple argument of timescales to determine whether an instability develops independently in one of the fluids or as a coherent instability in both. If the timescales of the instability growth in one of the fluids is fast compared to the collisional coupling then a decoupled instability (with the eigenvector of the instability dominated by terms for only one of the fluids) will occur. If the timescales are slow, then the instability-driven motions will be relatively similar in both fluids. In this regime, there will be a leading fluid (that is most unstable) and a following fluid (that is being dragged along). The leading and following fluids can be characterised by the magnitudes of the different terms in the eigenvector.

Soler et al. (2012) investigated the growth of the Kelvin–Helmholtz instability (an instability driven by shear flows) in a compressible two-fluid system. The growth rate ($\gamma_{KHI}$) found in the incompressible limit of their calculations for a range of velocity differences (the driver of the instability) is shown in the top panel of Fig. 6 where instabilities dominated by both neutral species modes and charged species modes can be seen. As the Lorentz force works to suppress instability in the plasma, naturally these modes are less unstable than the neutral-dominated modes. Three different levels of coupling were investigated, showing that the growth rate reduces for both neutral-dominated and charged species-dominated modes as coupling increases. For the neutral-dominated modes this is a sign of the drag of the charge-species slowing the growth of the instability to pull it closer to that of a single instability mode based on the stability properties of the bulk partially ionised plasma. For the charged species, as separate neutral-dominated and charged-species dominated modes do not exist in the fully coupled limit, the charged species-dominated modes have their growth rate decrease towards stability. Extending these ideas to a prominence thread, Martínez-Gómez et al. (2015) showed that even though magnetic fields might be able to stabilise the instability for a fully ionised plasma, the partial ionisation of the prominence plasma would allow for instability at observed flow speeds.

In a similar vein, the linear analysis of the Rayleigh–Taylor instability (an instability driven by baroclinicity, anti-alignment of density and pressure gradients, at a density inversion) in a partially ionised plasma has been investigated in both single-fluid (Díaz et al., 2014; Ruderman et al., 2018) and two-fluid frameworks (Díaz et al., 2012). The growth rate of this instability ($\gamma_{RTI}$) in the incompressible limit of the two-fluid instability against the magnitude of the driving buoyancy term is shown in the bottom panel of Fig. 6. As with the work of Soler et al. (2012), the Lorentz force is working to suppress the instability in the charged species-dominated modes, making the neutral-dominated modes the most unstable. Similarly, again, to

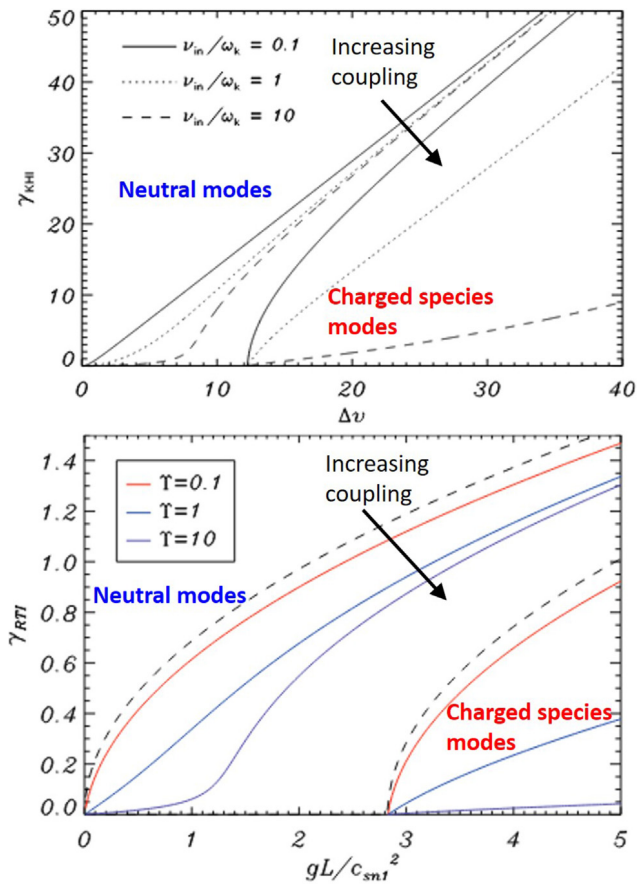

Fig. 6. Growth rates for the two-fluid Kelvin–Helmholtz instability (top panel, modified from Soler et al., 2012) and the two-fluid Rayleigh–Taylor instability (bottom panel, modified from Díaz et al., 2012) in the incompressible limit. In both cases larger values on the x-axis implies a larger driving force for the instability. Neutral-dominated and charged species-dominated modes are labelled. In each panel there are three different levels of coupling shown from weakest (solid line in top panel and red line in bottom), through intermediate (dotted line in top panel and blue line in bottom) to strongest (dashed line in top panel and purple line in bottom). The dashed lines in the bottom panel show the completely decoupled limit. In both panels the arrow shows how coupling results in a reduction of the growth rate. © AAS. Reproduced with permission.

Soler et al. (2012), as the coupling is increased the neutral-dominated mode morphs into the coupled-fluids mode and the isolated charged species-dominated mode tends towards stability.

The addition of compressibility in both the studies of Soler et al. (2012) and Díaz et al. (2012) leads to a further effect suppressing instability (energy has to be spent compressing the fluid instead of driving motions, e.g. Ruderman, 2017). This can switch the most unstable fluid from the neutral fluid to the plasma for these instabilities, or create situations where in a previously unstable system neither fluid is unstable at all, resulting in a complex picture of the stability (for example see Figure 6 of Soler et al., 2012).

### 4.2. Partial Ionisation in nonlinear instability dynamics

Linear stability theory is a powerful tool to understand whether a particular system can become unstable, however once that instability has grown and developed the subsequent behaviour is the realm of nonlinear theory. Due to the complexity of nonlinear dynamics we often require numerical simulations to make progress, the same is the case for instabilities in two-fluid partially ionised plasma models. For example, the work of Jones and Downes (2011) and Jones and Downes (2012) shows that for the Kelvin–Helmholtz instability in the nonlinear regime the decoupling of the bulk flow from the magnetic field results in a vast reduction in the magnetic energy of the nonlinear system. This instability has been observed to develop as a result of flows of prominence material (Berger et al., 2017; Yang et al., 2018; Hillier and Polito, 2018), as such merits understanding how partial ionisation will change its nonlinear behaviour.

The work of Hillier (2019) looks in detail at the nonlinear evolution of the Kelvin–Helmholtz instability in the case where multiple linear modes, spanning from coupled to decoupled, are excited in a system with both a velocity and density jump, and a stream-wise magnetic field. The natural response in this system is for neutral fluid to be unstable, but the magnetic field to suppress the development of the instability in the plasma leading to the drift of neutral material across the magnetic field. This can be seen clearly in Fig. 7 where the neutral vortices are moving the fluid across the magnetic field. These vortical structures were found to be where strong velocity drifts were present (and with this frictional heating). However, as the physical scale of the nonlinear instability layer became larger, the effective coupling became stronger reducing the velocity drift and with that the total heating from frictional heating in the layer (which for large layer widths scaled as the inverse of the layer width Hillier, 2019).

One of the key physical responses found in this system was in the thermal coupling. As neutral material moved across the magnetic field it interacted with plasma of a different temperature. When the ionisation fraction is small, the neutral fluid acts as a large source/sink of thermal energy, driving temperature changes in the local plasma. This leads to plasma pressure imbalances along the magnetic field, resulting in compressible motions developing as the plasma pressure works to create a pressure balance along the magnetic field. This turbulent transport of heat across the magnetic field by neutral drift has the potential to be very important when considering how prominence material or spicules interact with the coronal material that surrounds them, potentially aiding the mixing/cooling process proposed by Hillier and Arregui (2019).

Martínez-Gómez et al. (2021) studied the evolution of the Kelvin–Helmholtz in a partially ionised plasma, but with initially no magnetic field. They used Biermann battery term (e.g. Kulsrud and Zweibel, 2008) in induction equation which creates magnetic fields though the electron pressure and density distributions becoming out of alignment. In the simulations of Martínez-Gómez et al. (2021) Kelvin–Helmholtz vortices drove this term to create a magnetic field perpendicular to the plane of the simulation. For simulations where there was no collisional coupling between the fluids, there was (as would be expected) no variation in the results on the ionisation fraction. However, when the coupling terms between the fluids were switched on, it resulted in stronger magnetic fields being produced by a factor of ∼ 1.5.

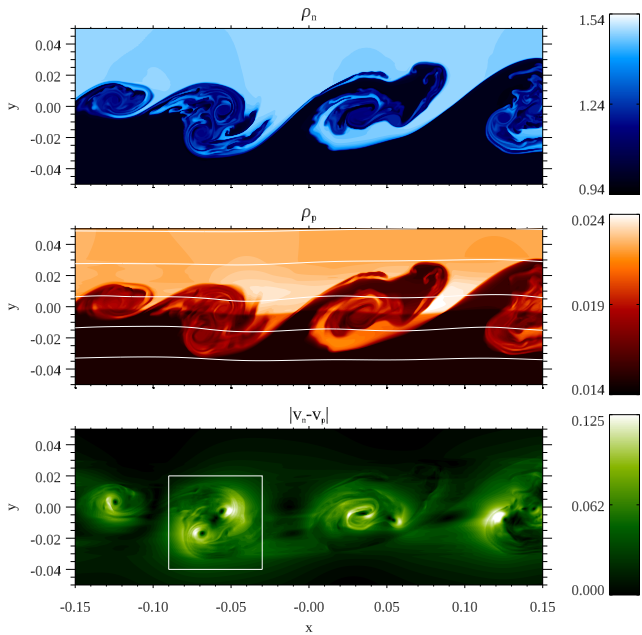

Fig. 7. Snapshot from Hillier, 2019 showing the neutral density (top panel), plasma density with magnetic field lines (middle panel) and magnitude of the velocity drift (bottom panel). Vortical structure in the density show the position of Kelvin–Helmholtz vortices. The total density was normalised to be 1 in the bottom layer and the velocity magnitude is normalised to the sound speed in that layer. Reproduced from Hillier, A. "Ion-neutral decoupling in the nonlinear Kelvin-Helmholtz instability: Case of field-aligned flow." Physics of Plasmas, 26(8), 082902, (2019)., with the permission of AIP Publishing.

Even for these relatively simple settings, there are still many fundamental instability driven physical process where the role of partial ionisation is still to be quantified. A key example of this would be the vortex disruption process investigated by Mak et al. (2017). In their study flow instabilities would wind up magnetic field until it became sufficiently stressed, underwent magnetic reconnection, ultimately destroying the vortex. How this behaviour will alter in partially ionised plasma, where we will see in Section 4.3 that partial ionisation can change reconnection dynamics, is still to be investigated.

### 4.2.1. Magnetic Rayleigh Taylor instability and partially ionised prominence dynamics

Moving beyond idealised models, we come to simulations of instabilities in partially ionised plasma that are set up to be directly relevant to understanding the dynamics of plasma of the solar atmosphere. In the studies of Braileanu et al. (2021a,b), the authors used two-fluid non-linear simulations to investigate the development of the Rayleigh–Taylor instability in a prominence thread to understand how two-fluid effects may be important in the development of plumes in solar prominences (for a review see Hillier, 2018). The work of Braileanu et al. (2021a,b) follows on from the study presented in Leake et al. (2014) and in some sense extending the study performed using the single fluid approximation by Khomenko et al. (2014b).

By seeding the Rayleigh–Taylor instability at a smooth boundary between the predominantly neutral prominence thread and the ionised corona below, it is natural that the gravitational potential energy of the neutral fluid is the driver of the instability in this case (Díaz et al., 2012). However, the actual linear stability of this system is quite complex. The fully-coupled limit (i.e. ideal MHD) was found to be unstable for any wavelength perturbation when there was no magnetic shear, with magnetic shear being necessary to create a cut-off wavelength (Braileanu et al., 2021a). However, the two-fluid simulations presented a cut-off wavelength below which the instability did not grow even without magnetic shear(Braileanu et al., 2021a). An explanation for this is that the initial conditions of both fluids (both out of equilibrium but held together through the collisional coupling) are baroclinically stable (i.e. the pressure and density gradients in each fluid are aligned). As it is the baroclinic term ($\nabla \rho \times \nabla P$) in the vorticity equation that drives the Rayleigh–Taylor instability (e.g Zhou et al., 2021), it is only when the coupling is sufficient to force the fluids to work together that they become a single baroclinically unstable fluid and can produce instability.

In the nonlinear stages of the instability, large velocity drifts of the order of $1\,\mathrm{km\,s^{-1}}$ between the neutral fluid and plasma developed (Braileanu et al., 2021a). These are drift velocities driven by both the natural separation of the fluids in the development of Rayleigh–Taylor plumes, and through the interaction of plumes especially in the case where magnetic shear means that current sheets develop as the plumes interact. By investigating how the magnitude of the coupling changed the nonlinear dynamics, they could show that lower coupling leads to more diffuse structures in the neutral fluid, because the fluid was able to slip more efficiently across the magnetic field (Braileanu et al., 2021b), though the stability of the individual fluids (as discussed above) may also have played an important role in this process as well.

This work was extended to include ionisation and recombination in Braileanu et al. (2021b). In Fig. 8 the left panels show the neutral and plasma densities when ionisation and recombination are included, the right when those terms are switched off in the equations. Though the density evolution of the neutral fluid is not noticeably effected, the plasma density is vastly different, with the downward plumes significantly more visible when ionisation and recombination allow the ambient plasma to respond to the neutrals as they drift across the magnetic field.

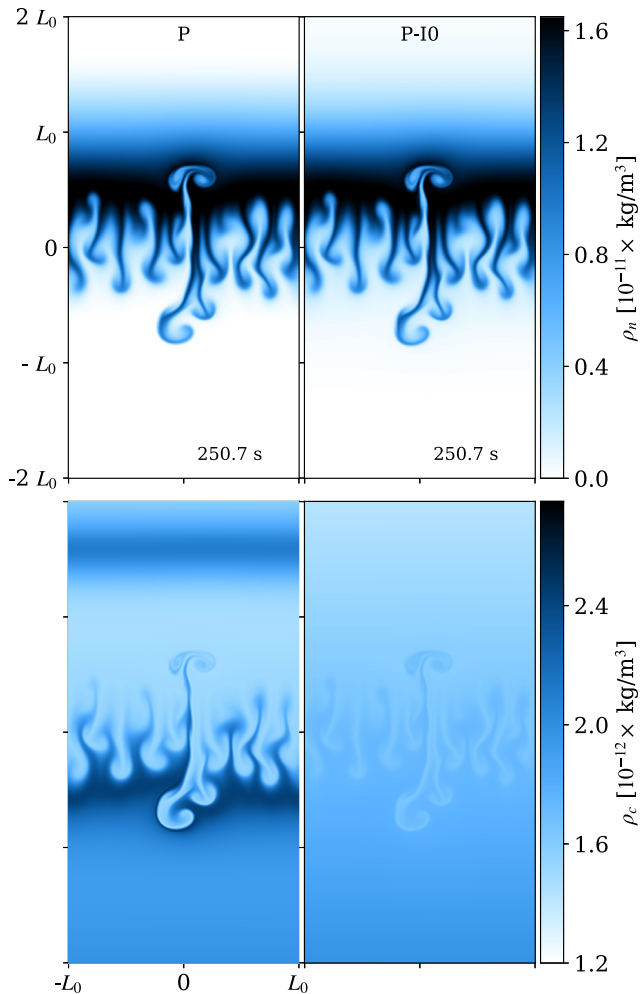

Fig. 8. Neutral (top) and plasma (bottom) densities for the case of Rayleigh–Taylor plumes at the boundary between a prominence thread and the ambient corona form Braileanu et al. (2021b). Differences between the case where ionisation and recombination are included (left) and neglected (right) are shown. Credit: Popescu Braileanu, B., Lukin, V. S., Khomenko, E. et al., A&A, 650, A181, 2021b, reproduced with permission © ESO.

### 4.3. Instabilities and magnetic reconnection

In terms of instabilities relating to magnetic reconnection there are two key instabilities to consider: the tearing instability (e.g. Zweibel, 1989) that results in the development of plasmoids in a current sheet, and the coalescence instability (e.g. Tajima and Sakai, 1986) that drives the newly formed plasmoids to interact. In combination these instabilities are seen as key for driving turbulence in a reconnecting current sheet, which ultimately results in fast magnetic reconnection (e.g. Loureiro et al., 2007; Loureiro et al., 2012). This process could be important for explaining the presence of fast, bursty reconnection observed in chromospheric plasma (Singh et al., 2011; Singh et al., 2012; Guo et al., 2020).

In a partially ionised plasma, as well as the natural collapse of gas-pressure supported current sheets (Brandenburg and Zweibel, 1994), the growth rate of the tearing mode can also greatly change. The analysis by Zweibel (1989) highlighted three regimes of the instability in a two-fluid model: the strongly coupled regime where both fluids evolve in the instability, the intermediate regime where the neutrals do not respond but exert a drag on the plasma and a decoupled regime where the instability grows in the plasma fluid. As these different regimes are associated with different levels of coupling, they can be connected with different scales of current sheets in the solar chromosphere (Singh et al., 2015). As a consequence, different physics controls the growth of plasmoids at different scales, modifying the potential fractal nature of plasmoids in a current sheet (Singh et al., 2015). See, for example, the review by Zweibel et al. (2011) for more details.

The development of plasmoids in a reconnecting current sheet in conditions relevant for chromospheric plasma has been studied through numerical simulations. As with fully ionised MHD simulations (e.g. Loureiro et al., 2012) once a sufficiently high Lundquist number (ratio of diffusion time to Alfvén time) is reached reconnecting current sheets in two-fluid partially ionised plasma simulations also become unstable to the formation of plasmoids (e.g. Leake et al., 2012; Ni et al., 2015). However, the change in instability physics with scale modifies the critical aspect ratio of a current sheet at which plasmoids can form (Pucci et al., 2020). The development of plasmoids in the current sheet produces highly time-dependent reconnection with average reconnection rates that are independent of the Lundquist number (Leake et al., 2012).

Once plasmoids have formed in the current sheet a second instability, the coalescence instability mentioned earlier, can play a crucial role in driving the interaction of neighbouring plasmoids. This instability is a current driven instability (developing as two currents flowing in the same direction attract each other). As the Lorentz force is the key driver of this instability it naturally develops on timescales of the order of the Alfvén time. Murtas et al. (2021) performed a detailed study of how the finite coupling between a plasma and neutral fluid changes the coalescence of plasmoids. As we can see from our discussion of Alfvén waves relating to Eq. 35 there are two Alfvén speeds, the bulk Alfvén speed and the plasma Alfvén speed, which results in two possible Alfvén times that could control the dynamics. The results of the parameter study by Murtas et al. (2021) show exactly this (see Fig. 9), weakly coupled systems coalescing on timescales determined by the plasma Alfvén time, and strongly coupled systems coalescing on timescales determined by the bulk Alfvén time. In between these two cluster points was an intermediate regime, covering a couple of decades of collision frequency, where the transition between coupled and decoupled occurs. the general consequence was that partially ionised plasma coalescence could occur on faster timescales than its fully ionised MHD counterpart. Related works by Smith and Sakai (2008) and Sakai and Smith (2009) focusing on plasmoid merging in the lower solar atmosphere to understand penumbral microjets (Katsukawa et al., 2007). Their results also show that plasmoid merger is sped up in a two-fluid system. Also they highlight that initial large neutral inflow velocities can trigger the strongest reconnection.

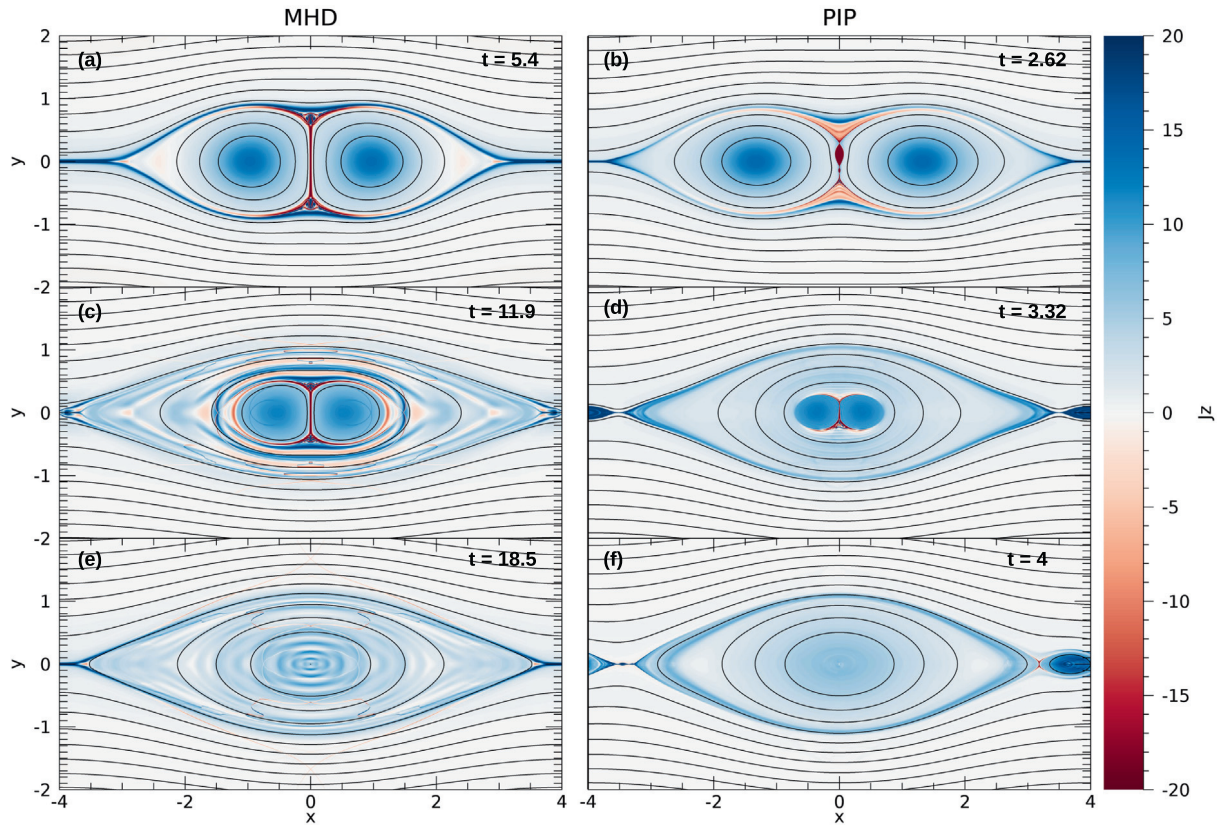

Fig. 9. Contour plots of field lines and current for the evolution of the coalescence instability in MHD (left column) and a two-fluid simulation (right column). Reproduced from Murtas, G., Hillier, A., and Snow, B. "Coalescence instability in chromospheric partially ionized plasmas." Physics ofPlasmas, 28(3), 032901, (2021), with the permission of AIP Publishing.

As plasmoids merge, the current sheet that develops between them can become unstable and form further plasmoids (as can be seen in Fig. 9). In general, it is expected that this subsequent plasmoid development is related to the timescale of the linear growth of the instability in the current sheet compared with the timescale for a plasmoid to be ejected (e.g. Pucci et al., 2020). This multiple plasmoid formation can result in the development of a fractal-like reconnection process (e.g. Singh et al., 2015), with the plasmoids potentially forming down to kinetic scales under the right conditions. In the simulations of Murtas et al. (2021) evidence was found of these tearing unstable plasmoids, i.e. the current sheet had reached a critical threshold and became unstable to the linear tearing instability, but also evidence of a separate plasmoid formation mechanism (called sub-critical plasmoid formation in their paper). In these simulations plasmoids were able to develop even though the current sheet did not reach any particular threshold for linear instability. In this case nonlinear two-fluid dynamics became important for driving secondary plasmoid formation, with the neutral drag changing the flow into the current sheet resulting in it pinching and forming a plasmoid as a result. Clearly the sum of all these effects implies that two-fluid physics can be very important for Chromospheric reconnection.

## 5. Discussion and looking towards the observations

In this review article we have looked at the general ideas behind using two-fluid modelling to investigate dynamics of the lower solar atmosphere. The key physical parameter that determines whether a fully ionised MHD or single-fluid partially ionised plasma or a two-fluid model (or on into kinetic modelling) is the most appropriate is the ratio of the dynamic frequency to the collision frequency. However, as has been argued here, many key dynamic phenomena of interest naturally scale multiple frequencies from those where single fluid models are applicable into the regimes where the physics becomes fundamentally two-fluid. Shocks, instabilities, and the nonlinear turbulence they create, are important examples where this occurs.

This paper has focused heavily on theoretical studies, and associated arguments and ideas, to understand the current state of our studies of the partially ionised solar atmosphere. The main reason behind this is it has been possible to make a lot of progress in developing theory and models, but conclusive observational studies have proved to be very difficult.

A number of attempts have been made to measure velocity differences between charged and neutral species in solar prominences. These seemed like the natural structure to make these observations as both the foreground and the background is the hot, tenuous, ionised corona which will not contribute to the emission in the cool spectral lines used to observe prominences. This then allows the Doppler velocity of different spectral lines to be measured, and by comparison of these velocities potentially identify if material of a different charge state is moving at a different velocity.

The results from these studies can be divided into two classes: those that claim an observation of ion-neutral drift (e.g. Khomenko et al., 2016; Wiehr et al., 2019; Wiehr et al., 2021) and those that see drift between all species (including between different neutral species, e.g. Anan et al., 2017). In the study of Khomenko et al. (2016) they measured velocity differences between neutral and ionised species up to a few $\mathrm{km\,s^{-1}}$. Interestingly this is a similar magnitude as found in the simulations of Braileanu et al. (2021a).

The explanation given in the studies that find Doppler velocity difference between even different neutral species is that along any line of sight in a prominence the relative distribution of charged and neutral species, and even excitation states of a given neutral species, is not uniform. Therefore, as different packets of material that form our line-of-sight view of the prominence move, the Doppler-shift observed in different spectral lines is different as it picks up the motion of different fluid packets and not because it is measuring the local drift between species (Anan et al., 2017). As this explanation can also explain the observations of drifts between neutral and charged species, it presents a challenge to really be able to identify genuine velocity drifts in prominences.

To remove these potential issues, Zapiór et al. (2022) applied strong constraint on the observed intensity of spectral lines based on radiative transfer calculations to remove any opacity effects in the calculation of prominence velocity drifts along with a number of spectral lines. This resulted in only a small area of the prominence being usable for the analysis, but with that area displaying systematic velocity differences of $\sim 1.7\,\mathrm{km\,s^{-1}}$ between charged and neutral species. As the systematic velocity differences between purely neutral species were one to two orders of magnitude smaller this could be the first categorical measurement of velocity differences between charges and neutral species in the solar atmosphere. It is interesting that the velocity difference is a systematic shift of $\sim 1.7\,\mathrm{km\,s^{-1}}$ as this is different to the velocity difference growing linearly with magnitude of the velocity of the prominence motions as found in Wiehr et al. (2019). Clearly more work, both observational and theoretical, is necessary to explain how these velocity drifts are developing.

Another interesting idea to measure velocity drifts was put forward by Anan et al. (2014), who proposed that the electric field felt by neutral particles as they drift across the magnetic field (i.e. the electric field given by $-\mathbf{v}_\mathrm{n} \times \mathbf{B}$) could be measurable through the Stark effect. Simulations of shocks suggest this field could be strong inside the shock front (Snow and Hillier, 2019). However, to measure this would require significant polarimetric accuracy. The design of the European Solar Telescope does seem it would likely be suited for performing these observations.

In the field of studying solar partially ionised plasma, it is our belief that the most important future progress will be made through better connection between theory and observations. Further development of observational studies, as discussed above, will be possible now with DKIST online and EST promises greater potential in the future. To make interpretation of the observations possible, effort is also needed to make two-fluid numerical models more realistic (in terms of the excitation and ionisation states of the particles) to allow for a closer one-to-one comparison with observations.

**Data availability**

There is no data produced for this manuscript.

**Declaration of Competing Interest**

The authors declare that they have no known competing financial interests or personal relationships that could have appeared to influence the work reported in this paper.

**Acknowledgements**

AH would like to thank Dr. S. Takasao and Dr. N. Nakamura for all their work in developing the (PIP) code and in discussing many ideas in partially ionised plasma. AH would also like to thank Prof. K. Shibata for giving him the freedom and encouragement that allowed the (PIP) code project to get started when it clearly was a distraction from other duties. Finally, AH and BS would like to thank Dr. G. Murtas for their work on magnetic reconnection using the (PIP) code which provided many insights for this review. AH and BS are supported by STFC Research Grant No. ST/R000891/1 and ST/V000659/1. AH was also supported by his STFC Ernest Rutherford Fellowship Grant No. ST/L00397X/1. AH would like to acknowledge the discussions with members of ISSI Team 457 "The Role of Partial Ionization in the Formation, Dynamics and Stability of Solar Prominences", which have helped improve the ideas in this manuscript. AH and BS would like to thnak the anonymous referees for their constructive comments. Hinode is a Japanese mission developed and launched by ISAS/JAXA, with NAOJ as domestic partner and NASA and STFC (UK) as international partners. It is operated by these agencies in co-operation with ESA and NSC (Norway).

**References**

Alharbi, A., Ballai, I., Fedun, V., Verth, G., 2022. Waves in weakly ionized solar plasmas. MNRAS 511, 5274–5286. https://doi.org/10.1093/mnras/stac444, arXiv:2202.07387.

Anan, T., Casini, R., Ichimoto, K., 2014. Diagnosis of Magnetic and Electric Fields of Chromospheric Jets through Spectropolarimetric Observations of H I Paschen Lines. ApJ 786, 94. https://doi.org/10.1088/0004-637X/786/2/94, arXiv:1402.4903.

Anan, T., Ichimoto, K., Hillier, A., 2017. Differences between Doppler velocities of ions and neutral atoms in a solar prominence. A&A 601, A103. https://doi.org/10.1051/0004-6361/201629979, arXiv:1703.02132.

Beckers, J.M., Tallant, P.E., 1969. Chromospheric Inhomogeneities in Sunspot Umbrae. Sol. Phys. 7, 351–365. https://doi.org/10.1007/BF00146140.

Berger, T., Hillier, A., Liu, W., 2017. Quiescent Prominence Dynamics Observed with the Hinode Solar Optical Telescope. II. Prominence Bubble Boundary Layer Characteristics and the Onset of a Coupled Kelvin-Helmholtz Rayleigh-Taylor Instability. ApJ 850, 60. https://doi.org/10.3847/1538-4357/aa95b6, arXiv:1707.05265.

Berger, T., Testa, P., Hillier, A., Boerner, P., Low, B.C., Shibata, K., Schrijver, C., Tarbell, T., Title, A., 2011. Magneto-thermal convection in solar prominences. Nature 472, 197–200. https://doi.org/10.1038/nature09925.

Berger, T.E., Shine, R.A., Slater, G.L., Tarbell, T.D., Title, A.M., Okamoto, T.J., Ichimoto, K., Katsukawa, Y., Suematsu, Y., Tsuneta, S., Lites, B.W., Shimizu, T., 2008. Hinode SOT Observations of Solar Quiescent Prominence Dynamics. ApJL 676, L89–L92. https://doi.org/10.1086/587171.

Berger, T.E., Slater, G., Hurlburt, N., Shine, R., Tarbell, T., Title, A., Lites, B.W., Okamoto, T.J., Ichimoto, K., Katsukawa, Y., Magara, T., Suematsu, Y., Shimizu, T., 2010. Quiescent Prominence Dynamics Observed with the Hinode Solar Optical Telescope. I. Turbulent Upflow Plumes. ApJ 716, 1288–1307. https://doi.org/10.1088/0004-637X/716/2/1288.

Brady, C.S., Arber, T.D., 2016. Simulations of Alfvén and Kink Wave Driving of the Solar Chromosphere: Efficient Heating and Spicule Launching. ApJ 829, 80. https://doi.org/10.3847/0004-637X/829/2/80, arXiv:1601.07835.

Braginskii, S.I., 1965. Transport Processes in a Plasma. Rev. Plasma Phys. 1, 276–284.

Brandenburg, A., Zweibel, E.G., 1994. The Formation of Sharp Structures by Ambipolar Diffusion. ApJL 427, L91–L94. https://doi.org/10.1086/187372.

Burkhart, B., Lazarian, A., Balsara, D., Meyer, C., Cho, J., 2015. Alfvénic Turbulence Beyond the Ambipolar Diffusion Scale. ApJ 805, 118. https://doi.org/10.1088/0004-637X/805/2/118, arXiv:1412.3452.

Carlsson, M., Stein, R.F., 2002. Dynamic Hydrogen Ionization. ApJ 572, 626–635. https://doi.org/10.1086/340293, arXiv:astro-ph/0202313.

Chae, J., Cho, K., Song, D., Litvinenko, Y.E., 2018. Nonlinear Effects in Three-minute Oscillations of the Solar Chromosphere. II. Measurement of Nonlinearity Parameters at Different Atmospheric Levels. ApJ 854, 127. https://doi.org/10.3847/1538-4357/aaa8e2.

Cheung, M.C.M., Cameron, R.H., 2012. Magnetohydrodynamics of the Weakly Ionized Solar Photosphere. ApJ 750, 6. https://doi.org/10.1088/0004-637X/750/1/6, arXiv:1202.1937.

Delmont, P., Keppens, R., 2011. Parameter regimes for slow, intermediate and fast MHD shocks. J. Plasma Phys. 77, 207–229. https://doi.org/10.1017/S0022377810000115.

Díaz, A.J., Khomenko, E., Collados, M., 2014. Rayleigh-Taylor instability in partially ionized compressible plasmas: One fluid approach. A&A 564, A97. https://doi.org/10.1051/0004-6361/201322147, arXiv:1401.5388.

Díaz, A.J., Soler, R., Ballester, J.L., 2012. Rayleigh-Taylor Instability in Partially Ionized Compressible Plasmas. ApJ 754, 41. https://doi.org/10.1088/0004-637X/754/1/41.

Draine, B.T., 1986. Multicomponent, reacting MHD flows. MNRAS 220, 133–148. https://doi.org/10.1093/mnras/220.1.133.

Draine, B.T., McKee, C.F., 1993. Theory of interstellar shocks. ARA&A 31, 373–432. https://doi.org/10.1146/annurev.aa.31.090193.002105.

Édel'Man, M.A., 1989. Corrugation instability of a strong parallel slow shock wave. I. Numerical calculations for the case of a radiative shock. Astrophysics 31, 656–663. https://doi.org/10.1007/BF01006842.

Gardner, C.S., Kruskal, M.D., 1964. Stability of Plane Magnetohydrodynamic Shocks. Phys. Fluids 7, 700–706. https://doi.org/10.1063/1.1711271.

Gary, G.A., 2001. Plasma Beta above a Solar Active Region: Rethinking the Paradigm. Sol. Phys. 203, 71–86. https://doi.org/10.1023/A:1012722021820.

Gilbert, H., Kilper, G., Alexander, D., 2007. Observational Evidence Supporting Cross-field Diffusion of Neutral Material in Solar Filaments. ApJ 671, 978–989. https://doi.org/10.1086/522884.

Gilbert, H.R., Hansteen, V.H., Holzer, T.E., 2002. Neutral Atom Diffusion in a Partially Ionized Prominence Plasma. ApJ 577, 464–474. https://doi.org/10.1086/342165.

Goedbloed, J.P., Keppens, R., Poedts, S., 2010. Advanced Magnetohydrodynamics: With Applications to Laboratory and Astrophysical Plasmas. Cambridge University Press, Cambridge. https://doi.org/10.1017/CBO9781139195560.

González-Morales, P.A., Khomenko, E., Vitas, N., Collados, M., 2020. Joint action of Hall and ambipolar effects in 3D magneto-convection simulations of the quiet Sun. I. Dissipation and generation of waves. A&A 642, A220. https://doi.org/10.1051/0004-6361/202037938, arXiv:2008.10429.

Grun, J., Stamper, J., Manka, C., Resnick, J., Burris, R., Crawford, J., Ripin, B.H., 1991. Instability of Taylor-Sedov blast waves propagating through a uniform gas. Phys. Rev. Lett 66, 2738–2741. https://doi.org/10.1103/PhysRevLett.66.2738.

Guidoni, S.E., Longcope, D.W., 2010. Shocks and Thermal Conduction Fronts in Retracting Reconnected Flux Tubes. ApJ 718, 1476–1490. https://doi.org/10.1088/0004-637X/718/2/1476, arXiv:1006.4398.

Guo, L.J., De Pontieu, B., Huang, Y.M., Peter, H., Bhattacharjee, A., 2020. Observations and Modeling of the Onset of Fast Reconnection in the Solar Transition Region. ApJ 901, 148. https://doi.org/10.3847/1538-4357/abb2a7, arXiv:2009.11475.

Hau, L.N., Sonnerup, B.U.Ö., 1989. On the structure of resistive MHD intermediate shocks. J. Geophys. Res. 94, 6539–6551. https://doi.org/10.1029/JA094iA06p06539.

Hillier, A., 2018. The magnetic Rayleigh-Taylor instability in solar prominences. Rev. Modern Plasma Phys. 2, 1. https://doi.org/10.1007/s41614-017-0013-2.

Hillier, A., 2019. Ion-neutral decoupling in the nonlinear Kelvin-Helmholtz instability: Case of field-aligned flow. Phys. Plasmas 26, 082902. https://doi.org/10.1063/1.5103248, arXiv:1907.12507.

Hillier, A., Arregui, I., 2019. Coronal Cooling as a Result of Mixing by the Nonlinear Kelvin-Helmholtz Instability. ApJ 885, 101. https://doi.org/10.3847/1538-4357/ab4795, arXiv:1909.11351.

Hillier, A., Matsumoto, T., Ichimoto, K., 2017. Investigating prominence turbulence with Hinode SOT Dopplergrams. A&A 597, A111. https://doi.org/10.1051/0004-6361/201527766, arXiv:1610.08281.

Hillier, A., Morton, R.J., Erdélyi, R., 2013. A Statistical Study of Transverse Oscillations in a Quiescent Prominence. ApJL 779, L16. https://doi.org/10.1088/2041-8205/779/2/L16, arXiv:1310.8009.

Hillier, A., Polito, V., 2018. Observations of the Kelvin-Helmholtz Instability Driven by Dynamic Motions in a Solar Prominence. ApJL 864, L10. https://doi.org/10.3847/2041-8213/aad9a5, arXiv:1808.02286.

Hillier, A., Polito, V., 2021. Observation of bi-directional jets in a prominence. A&A 651, A60. https://doi.org/10.1051/0004-6361/201935774.

Hillier, A., Shibata, K., Isobe, H., 2010. Evolution of the Kippenhahn-Schlüter Prominence Model Magnetic Field under Cowling Resistivity.

PASJ 62, 1231–1237. https://doi.org/10.1093/pasj/62.5.1231, arXiv:1007.1909.

Hillier, A., Takasao, S., Nakamura, N., 2016. The formation and evolution of reconnection-driven, slow-mode shocks in a partially ionised plasma. A&A 591, A112. https://doi.org/10.1051/0004-6361/201628215, arXiv:1602.01112.

Hollweg, J.V., Jackson, S., Galloway, D., 1982. Alfven Waves in the Solar Atmospheres - Part Three - Nonlinear Waves on Open Flux Tubes. Sol. Phys. 75, 35–61. https://doi.org/10.1007/BF00153458.

Houston, S.J., Jess, D.B., Asensio Ramos, A., Grant, S.D.T., Beck, C., Norton, A.A., Krishna Prasad, S., 2018. The Magnetic Response of the Solar Atmosphere to Umbral Flashes. ApJ 860, 28. https://doi.org/10.3847/1538-4357/aab366, arXiv:1803.00018.

Houston, S.J., Jess, D.B., Keppens, R., Stangalini, M., Keys, P.H., Grant, S.D.T., Jafarzadeh, S., McFetridge, L.M., Murabito, M., Ermolli, I., Giorgi, F., 2020. Magnetohydrodynamic Nonlinearities in Sunspot Atmospheres: Chromospheric Detections of Intermediate Shocks. ApJ 892, 49. https://doi.org/10.3847/1538-4357/ab7a90, arXiv:2002.12368.

Jara-Almonte, J., Ji, H., Yoo, J., Yamada, M., Fox, W., Daughton, W., 2019. Kinetic Simulations of Magnetic Reconnection in Partially Ionized Plasmas. Phys. Rev. Lett 122, 015101. https://doi.org/10.1103/PhysRevLett.122.015101.

Jones, A.C., Downes, T.P., 2011. The Kelvin-Helmholtz instability in weakly ionized plasmas: ambipolar-dominated and Hall-dominated flows. MNRAS 418, 390–400. https://doi.org/10.1111/j.1365-2966.2011.19491.x, arXiv:1107.4241.

Jones, A.C., Downes, T.P., 2012. The Kelvin-Helmholtz instability in weakly ionized plasmas - II. Multifluid effects in molecular clouds. MNRAS 420, 817–828. https://doi.org/10.1111/j.1365-2966.2011.20095.x, arXiv:1111.0436.

Katsukawa, Y., Berger, T.E., Ichimoto, K., Lites, B.W., Nagata, S., Shimizu, T., Shine, R.A., Suematsu, Y., Tarbell, T.D., Title, A.M., Tsuneta, S., 2007. Small-Scale Jetlike Features in Penumbral Chromospheres. Science 318, 1594–1597. https://doi.org/10.1126/science.1146046.

Khomenko, E., 2020. Multi-fluid extensions of mhd and their implications on waves and instabilities. In: MacTaggart, D., Hillier, A. (Eds.), Topics in Magnetohydrodynamic Topology, Reconnection and Stability Theory. Springer International Publishing, Cham, pp. 69–116. https://doi.org/10.1007/978-3-030-16343-3_3.

Khomenko, E., Cally, P.S., 2019. Fast-to-Alfvén Mode Conversion and Ambipolar Heating in Structured Media. II. Numerical Simulation. ApJ 883, 179. https://doi.org/10.3847/1538-4357/ab3d28.

Khomenko, E., Collados, M., Díaz, A., Vitas, N., 2014a. Fluid description of multi-component solar partially ionized plasma. Phys. Plasmas 21, 092901. https://doi.org/10.1063/1.4894106, arXiv:1408.1871.

Khomenko, E., Collados, M., Díaz, A.J., 2016. Observational Detection of Drift Velocity between Ionized and Neutral Species in Solar Prominences. ApJ 823, 132. https://doi.org/10.3847/0004-637X/823/2/132, arXiv:1604.01177.

Khomenko, E., Díaz, A., de Vicente, A., Collados, M., Luna, M., 2014b. Rayleigh-Taylor instability in prominences from numerical simulations including partial ionization effects. A&A 565, A45. https://doi.org/10.1051/0004-6361/201322918, arXiv:1403.4530.

Kulsrud, R.M., Zweibel, E.G., 2008. On the origin of cosmic magnetic fields. Rep. Prog. Phys. 71, 046901. https://doi.org/10.1088/0034-4885/71/4/046901.

Kuźma, B., Wójcik, D., Murawski, K., 2019. Heating of a Quiet Region of the Solar Chromosphere by Ion and Neutral Acoustic Waves. ApJ 878, 81. https://doi.org/10.3847/1538-4357/ab1b4a, arXiv:1906.01746.

Laming, J.M., Grun, J., 2002. Dynamical Overstability of Radiative Blast Waves: The Atomic Physics of Shock Stability. Phys. Rev. Lett 89, 125002. https://doi.org/10.1103/PhysRevLett.89.125002, arXiv:astro-ph/0207582.

Landau, L.D., Lifshitz, E.M., 1987. Fluid Mechanics, Second Edition: Volume 6 (Course of Theoretical Physics). Course of theoretical physics/ by L.D. Landau and E.M. Lifshitz, Vol. 6, 2nd ed., Butterworth-Heinemann, Oxford.

Leake, J.E., DeVore, C.R., Thayer, J.P., Burns, A.G., Crowley, G., Gilbert, H.R., Huba, J.D., Krall, J., Linton, M.G., Lukin, V.S., Wang, W., 2014. Ionized Plasma and Neutral Gas Coupling in the Sun's Chromosphere and Earth's Ionosphere/Thermosphere. Space Sci. Rev. 184, 107–172. https://doi.org/10.1007/s11214-014-0103-1, arXiv:1310.0405.

Leake, J.E., Lukin, V.S., Linton, M.G., Meier, E.T., 2012. Multi-fluid Simulations of Chromospheric Magnetic Reconnection in a Weakly Ionized Reacting Plasma. ApJ 760, 109. https://doi.org/10.1088/0004-637X/760/2/109, arXiv:1210.1807.

Leonardis, E., Chapman, S.C., Foullon, C., 2012. Turbulent Characteristics in the Intensity Fluctuations of a Solar Quiescent Prominence Observed by the Hinode Solar Optical Telescope. ApJ 745, 185. https://doi.org/10.1088/0004-637X/745/2/185, arXiv:1110.3159.

Lessen, M., Deshpande, N.V., 1967. Stability of magnetohydrodynamic shock waves. J. Plasma Phys. 1, 463–472. https://doi.org/10.1017/S0022377800003457.

Loureiro, N.F., Samtaney, R., Schekochihin, A.A., Uzdensky, D.A., 2012. Magnetic reconnection and stochastic plasmoid chains in high-Lundquist-number plasmas. Phys. Plasmas 19. https://doi.org/10.1063/1.3703318, 042303–042303. arXiv:1108.4040.

Loureiro, N.F., Schekochihin, A.A., Cowley, S.C., 2007. Instability of current sheets and formation of plasmoid chains. Phys. Plasmas 14. https://doi.org/10.1063/1.2783986, 100703–100703. arXiv:astro-ph/0703631.

Mak, J., Griffiths, S.D., Hughes, D.W., 2017. Vortex disruption by magnetohydrodynamic feedback. Phys. Rev. Fluids 2, 113701. https://doi.org/10.1103/PhysRevFluids.2.113701, arXiv:1609.03069.

Martínez-Gómez, D., Popescu Braileanu, B., Khomenko, E., Hunana, P., 2021. Simulations of the Biermann battery mechanism in two-fluid partially ionised plasmas. A&A 650, A123. https://doi.org/10.1051/0004-6361/202039113, arXiv:2104.06956.

Martínez-Gómez, D., Soler, R., Terradas, J., 2015. Onset of the Kelvin-Helmholtz instability in partially ionized magnetic flux tubes. A&A 578, A104. https://doi.org/10.1051/0004-6361/201525785, arXiv:1504.05379.

Meier, E.T., Shumlak, U., 2012. A general nonlinear fluid model for reacting plasma-neutral mixtures. Phys. Plasmas 19, 072508. https://doi.org/10.1063/1.4736975.

Morton, R.J., Verth, G., Hillier, A., Erdélyi, R., 2014. The Generation and Damping of Propagating MHD Kink Waves in the Solar Atmosphere. ApJ 784, 29. https://doi.org/10.1088/0004-637X/784/1/29, arXiv:1310.4650.

Murtas, G., Hillier, A., Snow, B., 2021. Coalescence instability in chromospheric partially ionized plasmas. Phys. Plasmas 28, 032901. https://doi.org/10.1063/5.0032236, arXiv:2102.01630.

Ni, L., Kliem, B., Lin, J., Wu, N., 2015. Fast Magnetic Reconnection in the Solar Chromosphere Mediated by the Plasmoid Instability. ApJ 799, 79. https://doi.org/10.1088/0004-637X/799/1/79, arXiv:1509.06895.

Niedziela, R., Murawski, K., Poedts, S., 2021. Chromospheric heating and generation of plasma outflows by impulsively generated two-fluid magnetoacoustic waves. A&A 652, A124. https://doi.org/10.1051/0004-6361/202141027, arXiv:2107.12050.

Nishizuka, N., Nakamura, T., Kawate, T., Singh, K.A.P., Shibata, K., 2011. Statistical Study of Chromospheric Anemone Jets Observed with Hinode/SOT. ApJ 731, 43. https://doi.org/10.1088/0004-637X/731/1/43.

Nishizuka, N., Shimizu, M., Nakamura, T., Otsuji, K., Okamoto, T.J., Katsukawa, Y., Shibata, K., 2008. Giant Chromospheric Anemone Jet Observed with Hinode and Comparison with Magnetohydrodynamic Simulations: Evidence of Propagating Alfvén Waves and Magnetic Reconnection. ApJL 683, L83–L86. https://doi.org/10.1086/591445, arXiv:0810.3384.

Nóbrega-Siverio, D., Martínez-Sykora, J., Moreno-Insertis, F., Carlsson, M., 2020a. Ambipolar diffusion in the Bifrost code. A&A 638, A79. https://doi.org/10.1051/0004-6361/202037809, arXiv:2004.11927.

Nóbrega-Siverio, D., Moreno-Insertis, F., Martínez-Sykora, J., Carlsson, M., Szydlarski, M., 2020b. Nonequilibrium ionization and ambipolar diffusion in solar magnetic flux emergence processes. A&A 633, A66. https://doi.org/10.1051/0004-6361/201936944, arXiv:1912.01015.

Okamoto, T.J., De Pontieu, B., 2011. Propagating Waves Along Spicules. ApJL 736, L24. https://doi.org/10.1088/2041-8205/736/2/L24, arXiv:1106.4270.

Okamoto, Takenori J., Liu, Wei, Tsuneta, Saku, 2016. Helical Motions of Fine-structure Prominence Threads Observed by Hinode and IRIS. ApJ 831 (2). https://doi.org/10.3847/0004-637X/831/2/126 126.

Okamoto, T.J., Tsuneta, S., Berger, T.E., Ichimoto, K., Katsukawa, Y., Lites, B.W., Nagata, S., Shibata, K., Shimizu, T., Shine, R.A., Suematsu, Y., Tarbell, T.D., Title, A.M., 2007. Coronal Transverse Magnetohydrodynamic Waves in a Solar Prominence. Science 318, 1577–1580. https://doi.org/10.1126/science.1145447, arXiv:0801.1958.

Pennicott, J.D., Cally, P.S., 2019. Smoothing of MHD Shocks in Mode Conversion. ApJL 881, L21. https://doi.org/10.3847/2041-8213/ab3423, arXiv:1907.10954.

Pereira, T.M.D., De Pontieu, B., Carlsson, M., Hansteen, V., Tarbell, T.D., Lemen, J., Title, A., Boerner, P., Hurlburt, N., Wülser, J.P., Martínez-Sykora, J., Kleint, L., Golub, L., McKillop, S., Reeves, K.K., Saar, S., Testa, P., Tian, H., Jaeggli, S., Kankelborg, C., 2014. An Interface Region Imaging Spectrograph First View on Solar Spicules. ApJL 792, L15. https://doi.org/10.1088/2041-8205/792/1/L15, arXiv:1407.6360.

Popescu Braileanu, B., Lukin, V.S., Khomenko, E., de Vicente, Á., 2019a. Two-fluid simulations of waves in the solar chromosphere. I. Numerical code verification. A&A 627, A25. https://doi.org/10.1051/0004-6361/201834154, arXiv:1905.03559.

Popescu Braileanu, B., Lukin, V.S., Khomenko, E., de Vicente, Á., 2019b. Two-fluid simulations of waves in the solar chromosphere. II. Propagation and damping of fast magneto-acoustic waves and shocks. A&A 630, A79. https://doi.org/10.1051/0004-6361/201935844, arXiv:1908.05262.

Popescu Braileanu, B., Lukin, V.S., Khomenko, E., de Vicente, Á., 2021a. Two-fluid simulations of Rayleigh-Taylor instability in a magnetized solar prominence thread. I. Effects of prominence magnetization and mass loading. A&A 646, A93. https://doi.org/10.1051/0004-6361/202039053, arXiv:2007.15984.

Popescu Braileanu, B., Lukin, V.S., Khomenko, E., de Vicente, Á., 2021b. Two-fluid simulations of Rayleigh-Taylor instability in a magnetized solar prominence thread. II. Effects of collisionality. A&A 650, A181. https://doi.org/10.1051/0004-6361/202140425, arXiv:2101.12731.

Pucci, F., Singh, K.A.P., Tenerani, A., Velli, M., 2020. Tearing Modes in Partially Ionized Astrophysical Plasma. ApJL 903, L19. https://doi.org/10.3847/2041-8213/abc0e7, arXiv:2006.03957.

Ruderman, M.S., 2017. Compressibility Effect on the Rayleigh-Taylor Instability with Sheared Magnetic Fields. Sol. Phys. 292, 47. https://doi.org/10.1007/s11207-017-1073-8.

Ruderman, M.S., Ballai, I., Khomenko, E., Collados, M., 2018. Rayleigh-Taylor instabilities with sheared magnetic fields in partially ionised plasmas. A&A 609, A23. https://doi.org/10.1051/0004-6361/201731534.

Sakai, J.I., Smith, P.D., 2009. Two-Fluid Simulations of Coalescing Penumbra Filaments Driven by Neutral-Hydrogen Flows. ApJL 691, L45–L48. https://doi.org/10.1088/0004-637X/691/1/L45.

Schunker, H., Cally, P.S., 2006. Magnetic field inclination and atmospheric oscillations above solar active regions. MNRAS 372, 551–564. https://doi.org/10.1111/j.1365-2966.2006.10855.x.

Singh, K.A.P., Hillier, A., Isobe, H., Shibata, K., 2015. Nonlinear instability and intermittent nature of magnetic reconnection in solar chromosphere. PASJ 67, 96. https://doi.org/10.1093/pasj/psv066, arXiv:1602.01999.

Singh, K.A.P., Isobe, H., Nishizuka, N., Nishida, K., Shibata, K., 2012. Multiple Plasma Ejections and Intermittent Nature of Magnetic Reconnection in Solar Chromospheric Anemone Jets. ApJ 759, 33. https://doi.org/10.1088/0004-637X/759/1/33.

Singh, K.A.P., Pucci, F., Tenerani, A., Shibata, K., Hillier, A., Velli, M., 2019. Dynamic Evolution of Current Sheets, Ideal Tearing, Plasmoid Formation and Generalized Fractal Reconnection Scaling Relations. ApJ 881, 52. https://doi.org/10.3847/1538-4357/ab2b99, arXiv:1904.00755.

Singh, K.A.P., Shibata, K., Nishizuka, N., Isobe, H., 2011. Chromospheric anemone jets and magnetic reconnection in partially ionized solar atmosphere. Phys. Plasmas 18. https://doi.org/10.1063/1.3655444, 111210–111210..

Smith, P.D., Sakai, J.I., 2008. Chromospheric magnetic reconnection: two-fluid simulations of coalescing current loops. A&A 486, 569–575. https://doi.org/10.1051/0004-6361:200809624, arXiv:0804.2086.

Snow, B., Hillier, A., 2019. Intermediate shock sub-structures within a slow-mode shock occurring in partially ionised plasma. A&A 626, A46. https://doi.org/10.1051/0004-6361/201935326, arXiv:1904.12518.

Snow, B., Hillier, A., 2020. Mode conversion of two-fluid shocks in a partially-ionised, isothermal, stratified atmosphere. A&A 637, A97. https://doi.org/10.1051/0004-6361/202037848, arXiv:2004.02550.

Snow, B., Hillier, A., 2021a. Collisional ionisation, recombination, and ionisation potential in two-fluid slow-mode shocks: Analytical and numerical results. A&A 645, A81. https://doi.org/10.1051/0004-6361/202039667, arXiv:2010.06303.

Snow, B., Hillier, A., 2021b. Stability of two-fluid partially ionized slow-mode shock fronts. MNRAS 506, 1334–1345. https://doi.org/10.1093/mnras/stab1672, arXiv:2106.04199.

Soler, R., Ballester, J.L., 2022. Theory of Fluid Instabilities in Partially Ionized Plasmas: An Overview. Front. Astron. Space Sci. 9, 789083. https://doi.org/10.3389/fspas.2022.789083.

Soler, R., Carbonell, M., Ballester, J.L., 2013a. Magnetoacoustic Waves in a Partially Ionized Two-fluid Plasma. ApJS 209, 16. https://doi.org/10.1088/0067-0049/209/1/16, arXiv:1309.7204.

Soler, R., Carbonell, M., Ballester, J.L., Terradas, J., 2013b. Alfvén Waves in a Partially Ionized Two-fluid Plasma. ApJ 767, 171. https://doi.org/10.1088/0004-637X/767/2/171, arXiv:1303.4297.

Soler, R., Díaz, A.J., Ballester, J.L., Goossens, M., 2012. Kelvin-Helmholtz Instability in Partially Ionized Compressible Plasmas. ApJ 749, 163. https://doi.org/10.1088/0004-637X/749/2/163, arXiv:1202.4274.

Stone, J.M., Edelman, M., 1995. The Corrugation Instability in Slow Magnetosonic Shock Waves. ApJ 454, 182–193. https://doi.org/10.1086/176476.

Suematsu, Y., Shibata, K., Neshikawa, T., Kitai, R., 1982. Numerical Hydrodynamics of the Jet Phenomena in the Solar Atmosphere - Part One - Spicules. Sol. Phys. 75, 99–118. https://doi.org/10.1007/BF00153464.

Tajima, T., Sakai, J.I., 1986. Explosive coalescence of magnetic islands. IEEE Trans. Plasma Sci. 14, 929–933. https://doi.org/10.1109/TPS.1986.4316643.

Terradas, J., Soler, R., Oliver, R., Ballester, J.L., 2015. On the Support of Neutrals Against Gravity in Solar Prominences. ApJL 802, L28. https://doi.org/10.1088/2041-8205/802/2/L28, arXiv:1503.05354.

Vranjes, J., Krstic, P.S., 2013. Collisions, magnetization, and transport coefficients in the lower solar atmosphere. A&A 554, A22. https://doi.org/10.1051/0004-6361/201220738, arXiv:1304.4010.

Wiehr, E., Stellmacher, G., Balthasar, H., Bianda, M., 2021. Velocity Difference of Ions and Neutrals in Solar Prominences. ApJ 920, 47. https://doi.org/10.3847/1538-4357/ac1791, arXiv:2108.13103.

Wiehr, E., Stellmacher, G., Bianda, M., 2019. Evidence for the Two-fluid Scenario in Solar Prominences. ApJ 873, 125. https://doi.org/10.3847/1538-4357/ab04a4, arXiv:1904.01536.

Wójcik, D., Kuźma, B., Murawski, K., Musielak, Z.E., 2020. Wave heating of the solar atmosphere without shocks. A&A 635, A28. https://doi.org/10.1051/0004-6361/201936938.

Yang, H., Xu, Z., Lim, E.-K., Kim, S., Cho, K.-S., Kim, Y.-H., Chae, J., Cho, K., Ji, K., 2018. Observation of the Kelvin-Helmholtz Instability in a Solar Prominence. ApJ 857, 115. https://doi.org/10.3847/1538-4357/aab789.

Zank, G.P., Adhikari, L., Zhao, L.L., Mostafavi, P., Zirnstein, E.J., McComas, D.J., 2018. The Pickup Ion-mediated Solar Wind. ApJ 869, 23. https://doi.org/10.3847/1538-4357/aaebfe.

Zapiór, M., Heinzel, P., Khomenko, E., 2022. Doppler-velocity Drifts Detected in a Solar Prominence. ApJ 934, 16. https://doi.org/10.3847/1538-4357/ac778a.

Zhang, F., Poedts, S., Lani, A., Kuźma, B., Murawski, K., 2021. Two-fluid Modeling of Acoustic Wave Propagation in Gravitationally

Stratified Isothermal Media. ApJ 911, 119. https://doi.org/10.3847/1538-4357/abe7e8, arXiv:2011.13469.

Zhou, Y., Williams, R.J.R., Ramaprabhu, P., Groom, M., Thornber, B., Hillier, A., Mostert, W., Rollin, B., Balachandar, S., Powell, P.D., Mahalov, A., Attal, N., 2021. Rayleigh-Taylor and Richtmyer-Meshkov instabilities: A journey through scales. Phys. D Nonlinear Phenomena 423, 132838. https://doi.org/10.1016/j.physd.2020.132838.

Zweibel, E.G., 1989. Magnetic Reconnection in Partially Ionized Gases. ApJ 340, 550–557. https://doi.org/10.1086/167416.

Zweibel, E.G., Lawrence, E., Yoo, J., Ji, H., Yamada, M., Malyshkin, L.M., 2011. Magnetic reconnection in partially ionized plasmas. Phys. Plasmas 18, 111211. https://doi.org/10.1063/1.3656960.